\documentclass[a4paper,12pt,preprint]{elsarticle}
\usepackage{amsmath,array,amssymb}
\usepackage{graphicx}

\usepackage{multicol}
\usepackage{color}
\usepackage{xspace}
\usepackage{enumerate}

\usepackage{hyperref}
\hypersetup{
  pdftitle={Opto-RF transduction in Er:CaWO4},
  pdfauthor={Thierry Chanelière}
}

\usepackage{braket}
\newcommand{\cmt}[1]{}

\renewcommand{\vec}[1]{\boldsymbol{#1}}

\newcommand{\eryso}[0]{{Er$^{3+}$:Y$_2$SiO$_5$}\xspace}

\newcommand{\yso}[0]{Y$_2$SiO$_5$\xspace}
\newcommand{\er}{Er$^{3+}$\xspace}

\newcommand{\cawo}[0]{CaWO$_4$\xspace}

\newcommand{\ercawo}[0]{{Er$^{3+}$:CaWO$_4$}\xspace}

\newcommand{\eT}{\exp\left(-\hbar \omega_s/kT\right)}

\usepackage{lineno}

\begin{document}

%\linenumbers

\begin{frontmatter}

\title{Opto-RF transduction in \ercawo}

\author{Thierry Chanelière\corref{cor1}}
\ead{thierry.chaneliere@neel.cnrs.fr}
\author{Rémi Dardaillon}
\author{Pierre Lemonde}
\author{Jérémie J. Viennot}
\address{Univ. Grenoble Alpes, CNRS, Grenoble INP, Institut N\'eel, 38000 Grenoble, France}

\author{Emmanuel Flurin}
\author{Patrice Bertet}
\address{Université Paris-Saclay, CEA, CNRS, SPEC, 91191 Gif-sur-Yvette Cedex, France}

\author{Diana Serrano }
\author{Philippe Goldner}
\address{Chimie ParisTech, PSL University, CNRS, Institut de Recherche de Chimie Paris, 75005 Paris, France}

\cortext[cor1]{Corresponding author}
 
%\date{\today}

%\maketitle

\begin{abstract}
We use an erbium doped \cawo crystal as a resonant transducer between the RF and optical domains at 12\,GHz and 1532\,nm respectively. We employ a RF resonator to enhance the spin coupling but keep a single-pass (non-resonant) optical setup. The overall efficiency is low but we carefully characterize the transduction process and show that the performance can be described by two different metrics that we define and distinguish: the  electro-optics and the quantum efficiencies. We reach an electro-optics efficiency of -84\,dB for 15.7\,dBm RF power. % corresponding to a V$_\pi$ of xxxV%
The corresponding quantum efficiency is -142\,dB for 0.4\,dBm optical power. We develop the Schr\"odinger-Maxwell formalism, well-known to describe light-matter interactions in atomic systems, in order to model the conversion process. We explicitly make the connection with the cavity quantum electrodynamics (cavity QED) approach that are generally used to describe quantum transduction.
\end{abstract}

\end{frontmatter}

\newpage
\tableofcontents

%\pacs{}

\newpage
\section{Introduction and basic principle}
In the bestiary of quantum technologies, the need has recently arisen for an optical to microwave transducer to bridge the gap between two frequency domains: quantum optical telecommunications traveling through fibre networks on the one hand and spin or superconducting qubits confined to very low temperature cryostats on the other, with no real possibility of interconnection. From a more fundamental point of view, the realisation of such a device would allow the light-matter interaction to be pushed to the quantum scale, both in the radio-frequency (RF) domain, which can be driven by electric, magnetic or acoustic fields, and the optical domain in the same material system \cite{Andrews2014, Vainsencher, PhysRevApplied.14.061001, Mirhosseini2020,  Delaney2022, PRXQuantum.1.020315, Sahu2022, Sahu2023, arnold2023alloptical}. These questions and a unified presentation are well covered by several review articles \cite{Lambert, Lauk_2020, Han:21}.
Among the many possible materials, rare earth ions doped crystals were the first to be considered historically \cite{PhysRevLett.113.063603, PhysRevLett.113.203601, PhysRevA.99.063830}, since they have been studied during decades in parallel by optical spectroscopy and electron spin resonance (ESR), enabling a detailed understanding of properties for applications in magnetism or laser development.

The key parameters that drive the efficiency of the transduction process are the optical and spin cooperativities, and more precisely the product of the two quantities. In practice, lightly doped samples as the ones historically considered \cite{10.1063/1.1674609, Faure1996-ep}, few tens of ppm in our case, already offer a significant single-pass absorption. Indeed, the optical depth approaches one in many crystals. This quantity actually defines the optical cooperativity in free space propagation, meaning without cavity. This experimental fact stimulated the development of optical quantum memories in solid-state atomic ensembles \cite{CHANELIERE201877}. Concerning the RF cooperativity, a resonator is necessary to enhance the spin-microwave interaction and probe diluted samples. Starting from standard ESR apparatus with simple rectangular cavities in historical experiments \cite{antipin1968paramagnetic, MASON1968260, mims1968phase}, modern ESR studies now employ superconducting circuits \cite{mr-1-315-2020}.

That being said, the dual integration of RF and photonic circuits is very beneficial for enhancing the light-matter interactions. This approach has already made it possible to increase transduction efficiency \cite{rochman2023microwave, bartholomew2020chip}. These developments are particularly timely, as they can be carried out in synergy with current efforts to integrate quantum memories, by developing waveguides and nanophotonic devices \cite{Photonic_Integrated_QM}. Despite the tour de force that these achievements represent, they remain difficult to characterize, precisely because of their integration. For example, in a doubly resonant optical and RF scheme, it remains difficult to extract the ion couplings and the passive resonator characteristics independently.

Based on this observation, we have then decided to operate a simplified setup to observe opto-RF transduction in \ercawo, employing a rectangular RF resonator to drive the spins and a single-pass laser beam to excite the optical transition. This kind of setup has been widely employed for Raman heterodyne detection, even recently \cite{PhysRevA.92.062313}. The initial goal of this detection method is to perform an optical measurement of the spin transitions without quantitative analysis of the emitted signal, as it was the case in the early days of the technique \cite{PhysRevLett.50.993, PhysRevB.31.6947}.
The model developed at the time, although aimed at the qualitative analysis of nuclear magnetic resonance spectra with low optical absorptions, remains perfectly relevant and applicable to electron spins \cite{PhysRevB.28.4993}. We here derive an equivalent Schr\"odinger-Maxwell model for the equations of motion. Our work can alternatively be seen as the quantitative extension of the Raman heterodyne method since we carefully model the intensity of the beating signal that defines the transduction efficiency. It should also be noted that this is the first time the mixing process is observed in \ercawo.

\cawo is an interesting and widely studied material, which can be produced by a variety of growth techniques \cite{growth1,growth2}, because of its relatively low melting temperature compared to the iconic \yso, which continues to be the talk of the town in rare-earth ions based quantum technologies \cite{kunkel}. The \cawo matrix has a low nuclear spin density \cite{Ferrenti2020-xh}. This allows the observation of remarkably long spin coherence times \cite{bertaina2007rare, dantec}. This renewed effort has enabled ultimate demonstrations of both optical \cite{Ourari2023-vc} and RF single-ion detections \cite{Billaud_PhysRevLett.131.100804, wang:hal-03960036}.

\begin{figure}[ht]
\centering
\includegraphics[width=\columnwidth]{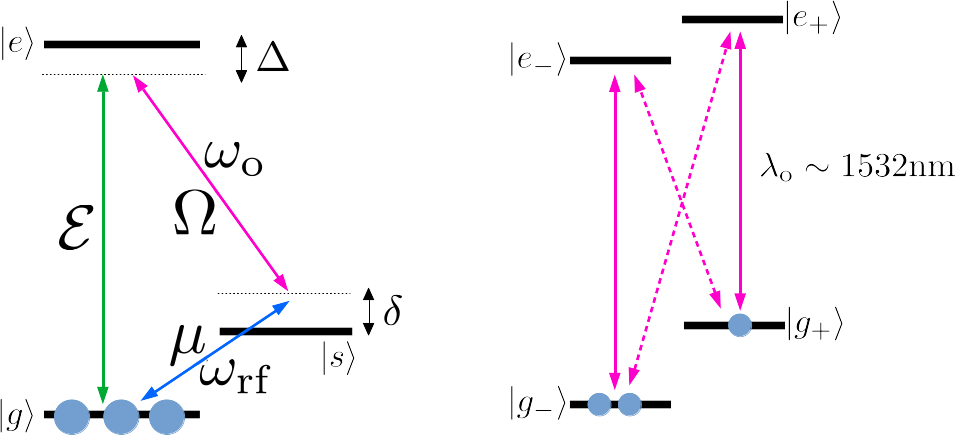} 
\caption{Left: Ideal $\Lambda$-scheme to implement resonant transduction. A low level spin transition, at few GHz between $\ket{g}$ and $\ket{s}$, is excited by the microwave field at $\omega_\mathrm{rf}$. The laser excites the $\ket{s} - \ket{e}$ transition in the optical domain at $\omega_\mathrm{o}$. This interaction are resonant or nearly resonant (detuned by few linewidths). The transduction signal (green arrow) is generated at $\omega_\mathrm{o}+\omega_\mathrm{rf}$. The field are characterized by their Rabi frequencies $\mu$, $\Omega$ and $\mathcal{E}$ for the optical Raman, microwave and transduction signal fields respectively. $\Delta$ and $\delta$ are the field detunings from the optical and spin transitions respectively.
Right: Realistic level structure of \ercawo. The ground and optically excited states are actually spin Kramers doublets that split under magnetic field into $\ket{g_-}$,$\ket{g_+}$ and $\ket{e_-}$,$\ket{e_+}$ respectively. Additionally, in our temperature range ($\sim 2-3$K), the spins are only partially polarized so both $\ket{g_-}$ and $\ket{g_+}$ are populated. As a consequence, the transduction is possible on four transitions between the two doublets. This will be discussed in more details in sections \ref{sec:absorption} and \ref{sec:wavelength}}
\label{fig:levels}
%/home/thierry/neel_ownCloud/neel_exchange/20220505_ErCaWO4/
\end{figure}

Before detailing the experimental setup, we first recall the basic principle of transduction using resonant interactions of the spin and optical transitions in a so-called $\Lambda$-scheme as summarized in Fig.\ref{fig:levels} (left). A more realistic structure of \er in \cawo under magnetic field is represented by a four-level structure in Fig.\ref{fig:levels} (right).

After having presented the experimental setup in section \ref{sec:setup}, we will then characterize the cooperativities of the optical and spin transitions independently in  section \ref{sec:cooperativities}. Finally, we present the results of the transduction process and its variation as a function of the accessible experimental parameters in section \ref{sec:transduction}. We continue with a detailed quantitative modeling of the transduction signal intensity in section \ref{sec:model}, and compare these theoretical results with the experiment. In the final section, we will discuss the discrepancy between the prediction and the experiment, and suggest some avenues for further studies in section \ref{sec:analysis}.

\section{Experimental setup}\label{sec:setup}

\subsection{Optical and RF excitations}
We place the \ercawo crystal ($5\times4\times3$\,mm) in a rectangular RF copper resonator (dimensions $a\times b \times d=15\times10\times20$\,mm parallel to our reference frame $\vec{x},\vec{y},\vec{z}$ respectively ) inside the inner bore of a superconducting coil providing the bias magnetic field $\vec{B_0}$, typically 100\,mT, along $\vec{z}$ in our case (see Fig.\ref{fig:detection}). The 4-mm dimension of the crystal along $\vec{y}$ corresponds to the laser propagation parallel to the crystalline c-axis. The nominal concentration of erbium is 50\,ppm (natural abundance). We excite the spins with a vector network analyzer (VNA) at $\sim$12\,GHz (Anritsu MS46522B) through a single coaxial line and collect the reflexion $S_{11}$ coefficient. This allows measurement of the RF cooperativity. The exact position of the crystal in the resonator and the intracavity field profile are detailed in \ref{appendix:AC_field}.

%%\begin{figure}[ht]
%%\centering
%%\includegraphics[width=.5\columnwidth]{xstal_resonator.png} 
%%\caption{The crystal is placed in a rectangular copper resonator in the inner bore of a bias coil at cryogenic temperature ($\sim 2.5$K). The laser resonant with the optical transition passes though the crystal (single-pass along its c-axis) by a 1mm diameter aperture in the cooper box. The RF excitation is provided by the pin of a SMA connector (antenna). $\vec{B_0}$ is the bias magnetic field.}
%%\label{fig:xstal_resonator}
%%/home/thierry/neel_ownCloud/neel_exchange/20220505_ErCaWO4/
%%\end{figure}

Following \cite[eq.(6.40)]{pozar1990microwave}, we expect the lowest resonant frequency to appear at $12.49$\,GHz. In practice, we observe a narrow resonance near $12.29$\,GHz with a quality factor of few thousands depending on the operating conditions (temperature, position of the crystal and length of the antenna). This resonance will be used for the spin excitation in the following. The length and position of the antenna (see Fig.\ref{fig:detection}) are adjusted at room temperature to approach the critical coupling to the feed line so $S_{11}$ goes to zero in order to maximally excite the spins for a given input power.

The laser beam from a Velocity\textsuperscript{\textregistered} External Cavity Diode Laser around $1532$\,nm enters and leaves the copper box through two holes (1mm diameter), crossing the crystal  in a single-pass along its c-axis (4-mm dimension, along $\vec{y}$).

\subsection{Transduction detection} \label{transduction_detection}

\begin{figure}[ht]
\centering
\includegraphics[width=\columnwidth]{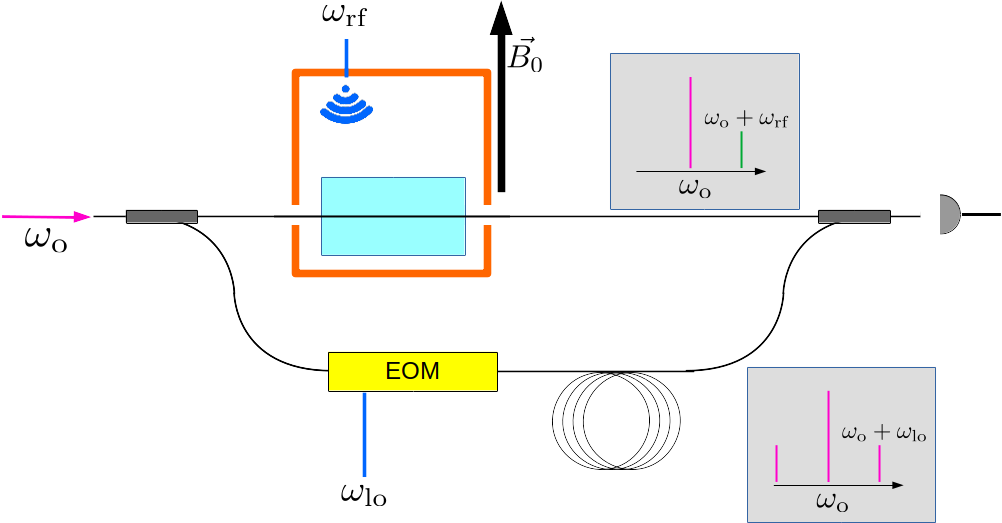} 
\caption{Transduction measurement setup. Half of the laser power, resonant with the optical transition passes through the crystal (single-pass along $\vec{y}$, parallel to the crystalline c-axis) by a 1-mm diameter aperture in the cooper box (upper arm). The crystal is placed in a rectangular copper resonator in the inner bore of a bias coil cooled down at cryogenic temperature ($\sim 2.5$\,K).  The RF excitation is provided by the pin of a SMA connector (antenna). $\vec{B_0}$ is the bias magnetic field along $\vec{z}$.
The heterodyne detection beats at $\omega_\mathrm{rf}-\omega_\mathrm{lo}$ between the opto-RF signal (upper arm) and the local oscillator generated by a fiber EOM (lower arm). The transduction signal appears as a sideband at $\omega_\mathrm{o}+\omega_\mathrm{rf}$ (see upper insert) after passing through the crystal. This beats on the photodiode with the  $\omega_\mathrm{o}+\omega_\mathrm{lo}$ component of the local oscillator generated by the EOM (see lower insert). We adjust the beam polarization by using a fiber paddle polarization controller after the crystal (before recombination with the LO) to maximize the interference contrast on the photodiode. Both frequencies $\omega_\mathrm{rf}$ and $\omega_\mathrm{lo}$ are delivered by a dual channel VNA.}
\label{fig:detection}
%/home/thierry/neel_ownCloud/neel_exchange/20220505_ErCaWO4/
\end{figure}

The transduction signal should appear as a $\omega_\mathrm{rf}= 2 \pi \times 12.29$\,GHz modulation of the transmitted laser beam at $\omega_\mathrm{o}$. This represents a relatively large frequency for optoelectronics devices so we decided to perform an heterodyne detection by generating a local oscillator (LO) $\omega_\mathrm{lo}= 2 \pi \times (  12.29$\,GHz$+44$\,MHz) from the same laser using a standard lithium niobate electro-optic frequency modulator (EOM, see Fig.\ref{fig:detection}). A low bandwidth photodiode ($<50$\,MHz of a Thorlabs PDA255) can then record the beating between $\omega_\mathrm{o}+\omega_\mathrm{rf}$ and  $\omega_\mathrm{o}+\omega_\mathrm{lo}$ at $44$\,MHz in our case. The choice of this beatnote frequency is justified in  \ref{appendix:choice_beating}.

\subsection{Transduction efficiency calibration}\label{sec:calibration}

\subsubsection{Electro-optics efficiency}

Before considering the quantum perspectives of the transducer, we propose to use a classical metric to evaluate the opto-RF conversion efficiency. We here define the electro-optics efficiency as the intensity of the sideband shifted by $\omega_\mathrm{rf}$ with respect to the total intensity. This is a simple classical characterization of the transduction process that is directly related to experimentally measurable parameters.

The raw amplitude of the beatnote at $\omega_\mathrm{rf}-\omega_\mathrm{lo}$ depends on the relative laser power between the LO and the crystal transducer arms that form a Mach-Zender interferometer after recombination in Fig.\ref{fig:detection}. On the LO side, the fiber EOM efficiency written $\eta_\mathrm{lo}$ (as the normalized intensity of the $\omega_\mathrm{lo}$ sideband) scales the amplitude of the recorded beatnote at $\omega_\mathrm{rf}-\omega_\mathrm{lo}$. We then measure  $\eta_\mathrm{lo}$ independently by comparing the amplitude at $\omega_\mathrm{lo}$ with respect to the carrier through a commercial Fabry-Perot interferometer. At our working frequency $\sim$12\,GHz, the fiber EOM efficiency is only $\eta_\mathrm{lo} = 6.8$\%$\pm0.3$\% at 18.4\,dBm input power (maximum power delivered by the VNA) because we exceed the nominal modulation bandwidth of the EOM and have a limited input driving power. Still, this is sufficient to observe and characterize the transduction signal.

Finally, there is no need to measure the power on both arms of the Mach-Zender interferometer because the transduction signal is a weak modulation of the carrier. Instead, we simply observe the Mach-Zender interferometer output DC signal on the photodiode. This fluctuates because the phase of the few meters long interferometer is not stable in time. There is no need to stabilize it for an AC measurement anyway. We record the amplitude of these slow temporal fluctuations between a maximum and a minimum as the voltage $V_\mathrm{fringes}$ that reveals the beating contrast (typically 80\% in our case). This latter may be imperfect because of a small polarization mismatch that we could not compensate by the fiber paddle polarization controller that we insert before the beams recombination (see Fig.\ref{fig:detection}). The spatial mode overlap is ensured by the single mode fiber recombiner. At the end, if the beatnote amplitude voltage recorded by the spectrum analyzer is $V_\mathrm{sa}$ then the electro-optics transduction efficiency $\eta_\mathrm{eo}$ is

\begin{equation}\label{eq:eom_eff}
\eta_\mathrm{eo}=\frac{1}{\eta_\mathrm{lo}} \left(\frac{V_\mathrm{sa}} {V_\mathrm{fringes}} \right)^2 
\end{equation}
This formula compares the measured power in the transduction sideband $\displaystyle \left(V_\mathrm{sa} \right)^2$ and the power in the local oscillator sideband $\displaystyle \eta_\mathrm{lo} \left(V_\mathrm{fringes}\right)^2$ that interfere to produce the beating signal as recorded independently. We systematically calibrate the spectrum analyzer so that we can compare the two amplitudes recorded by the spectrum analyzer and the oscilloscope, respectively.

We will use the electro-optics efficiency $\eta_\mathrm{eo}$ (expressed in dB typically) as a metric to analyze the experimental results. Nevertheless, it is important to carefully define the quantum number efficiency in the recent stimulating context.

\subsubsection{Quantum number efficiency}
By definition, the quantum number efficiency is the main metric for the quantum conversion process. As opposed to the electro-optics efficiency, we do not compare the intensity of the sideband at $\omega_\mathrm{rf}$ with respect to the optical carrier. Instead we here compare the photon number flux in the optical sideband at $\omega_\mathrm{o}+\omega_\mathrm{rf}$ and the RF photon flux. The powers are then $ \eta_\mathrm{eo} P_\mathrm{o}$ for the optical sideband and $P_\mathrm{rf}$ for the RF respectively, where $P_\mathrm{o}$ is the laser power in the crystal and  $P_\mathrm{rf}$ the RF power at the resonator input. So the quantum efficiency $\eta_\mathrm{Q}$ reads as 

\begin{equation}\label{eq:eom_Q}
\eta_\mathrm{Q}= \frac{\left( \eta_\mathrm{eo} P_\mathrm{o} \right)/\omega_\mathrm{o}} {P_\mathrm{rf}/\omega_\mathrm{rf}} =\eta_\mathrm{eo} \frac{  P_\mathrm{o} \omega_\mathrm{rf}} {P_\mathrm{rf} \omega_\mathrm{o}} 
\end{equation}

Eqs. \eqref{eq:eom_eff} and \eqref{eq:eom_Q} connect the two definitions of the efficiencies. 

The powers $P_\mathrm{o}$ and $P_\mathrm{rf}$ are experimental parameters. They are usually taken at the device inputs to focus on the conversion process and compensate for parasitic losses than can be avoided by technical improvements. Anticipating the description in section \ref{sec:spin_detuning}, at the maximum RF source power, we have $P_\mathrm{rf} = 37$\,mW. The laser power inside the cryostat (corrected from losses though cryostat windows and crystal interfaces) is $P_\mathrm{o} = 1.05$\,mW. The power ratio (maximum RF) is then $\displaystyle \frac{P_\mathrm{o}} {P_\mathrm{rf} } = 2.8 \times 10^{-2}$ or -15.5\,dB in log scale. Even if the powers have different orders of magnitude, we haven't observed any saturation of both transitions confirming that the efficiencies are constant and well-defined in our range of parameters. Although the saturation intensity of the optical transitions is generally low for the rare-earth ions, we use an almost collimated beam with a diameter of 800\,$\mu$m to avoid saturation. 

The frequency ratio is fixed by the level structure, in our case we have $\displaystyle \frac{\omega_\mathrm{rf}} {\omega_\mathrm{o}} = 6.3 \times 10^{-5}$ or -42.0\,dB in log scale.

In typical experimental conditions (maximum power), we have $\eta_\mathrm{Q}=\eta_\mathrm{eo}-57.5$\,dB when expressed in log scale. In conclusion, the efficiencies are proportional, they both characterize the transduction process that will be detailed in the next section.

\section{Transduction}

As discussed in the introduction, the RF and optical cooperativities are the key parameters to evaluate the transduction efficiency. They will be first characterized.

The reference method to measure the spin cooperativity consists in monitoring the $S_{11}$ from the VNA as the transition is tuned on resonance by sweeping the magnetic field. The shift induced by the spin scales as the RF cooperativity as we will show in section \ref{sec:RF cooperativity}.

Concerning the optical cooperativity, since we employ a non-resonant setup, the single-pass optical transmission can be directly recorded by a photodiode. As shown in section \ref{sec:absorption}, the cooperativity, equal to the optical depth in free space, can be deduced from the absorption spectrum.

We will then present the results of the transduction experiment in section \ref{sec:transduction}.

\subsection{Cooperativities}\label{sec:cooperativities}
\subsubsection{RF cooperativity}\label{sec:RF cooperativity}

\begin{figure}[ht!]
\centering
\includegraphics[width=\columnwidth]{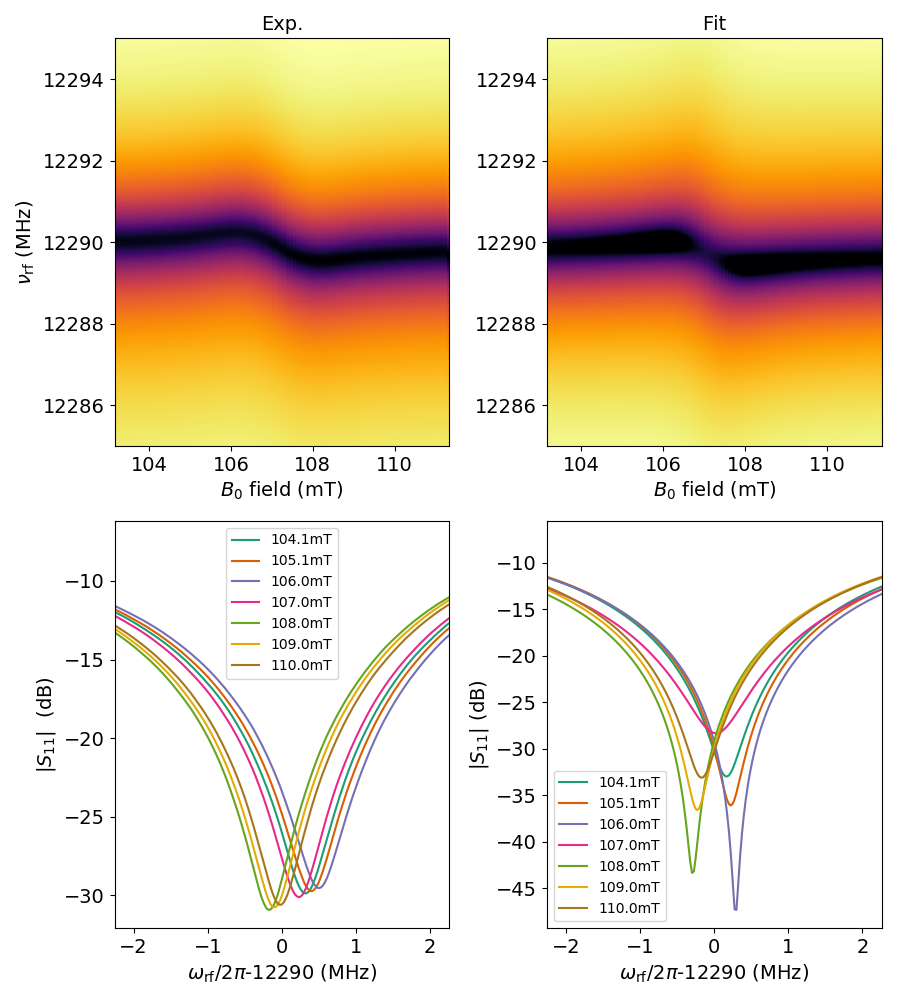} 
\caption{Absolute value of the $S_{11}$ coefficient. Top: Map (log. scale) of $|S_{11}|$ as a function of RF frequency (Y-axis) for a varying magnetic field (X-axis). Bottom: $|S_{11}|$ spectra for different characteristic values of the  magnetic field. The experimental results are on the left and corresponding fit on the right (see the text for details). We observe a cavity resonance shift when the spin resonance is swept across.}
\label{fig:cooperativity}
\end{figure}

The RF cooperativity can be deduced from the $S_{11}$ spectra for a varying magnetic field when the spin resonance is swept across as shown in Fig. \ref{fig:cooperativity}.

The cavity response is described by the well-established input-output theory \cite{Diniz_PhysRevA.84.063810, Julsgaard_PhysRevA.85.013844}
\cite[Eq.(3.50)]{ledantec:tel-03579857}
\begin{equation}
S_{11}(\omega_\mathrm{rf})=1-\frac{i \kappa_c}{\omega_\mathrm{rf}-\omega_c + i\frac{\kappa_t}{2}-W(\omega_\mathrm{rf})} \label{eq:S11}
\end{equation}
where $\kappa_c$ is the cavity coupling rate and $\kappa_t$ the damping rate ($\omega_c$ is the cavity resonant frequency). The spin interaction is contained in $W(\omega_\mathrm{rf})$
\cite[Eq.(3.53)]{ledantec:tel-03579857}

\begin{equation}
W(\omega_\mathrm{rf})= C_\mu \frac{i\kappa_t/2}{2i\frac{\omega_\mathrm{rf}-\omega_s}{\gamma} -1} \label{eq:W}
\end{equation}

where $C_\mu$ is RF cooperativity \cite{Julsgaard_PhysRevA.85.013844}, $\gamma$ the spin linewidth (FWHM) and $\omega_s$ the spin resonant frequency that varies with the magnetic field $B_0$ as $\omega_s=g \mu_B B_0 /\hbar$ ($g$ is the erbium g-factor along $\vec{B_0}$ and $\mu_B$ the Bohr magneton).

We fit the model given by Eqs. \eqref{eq:S11} \& \eqref{eq:W} to the map of $|S_{11}|$ in Fig. \ref{fig:cooperativity}. So we obtain for the different parameters:
$\kappa_t=2\pi\times(8.57 \pm 0.05)$\,MHz corresponding to a quality factor of 1430, $\kappa_c=2\pi\times(4.52 \pm 0.03)$\,MHz demonstrating a close to critical coupling condition with $\kappa_t \approx 2\kappa_c$. The spin linewidth is measured to be $\gamma=2\pi\times(219.14 \pm 0.09)$\,MHz limited by the magnetic field inhomogeneity along the crystal dimensions. This value corresponds to a 1.7\% magnetic field inhomogeneity. This is a typical value for a coil of this dimension (one-inch inner bore). The crystal may be also off-centered in the solenoid further increasing the broadening by moving away from the field maximum at the central axis. 
Additionally the fitted RF cooperativity is $C_\mu=0.135034\pm 0.000002$ in satisfying agreement with spin concentration and sample size (see \ref{appendix:spin_cooperativity} for details). The results also gives an erbium g-factor of $g=8.20\pm 0.04$ in correct agreement with the reported value of $8.45$ using this field orientation for $g_\perp$ \cite{enrique_optical_1971, bertaina2007rare}. 
We finally extract a total attenuation coefficient of $-5.4$\,dB mostly attributed to the coaxial line inside the cryostat.

By looking at Fig. \ref{fig:cooperativity}, we note that the fit predicts lower values for $|S_{11}|$ reaching -45\,dB as compared to the minimum -30\,dB that we observe. An ad-hoc introduction of extra losses (typically -15\,dB) would certainly improve the fit accuracy, but this is difficult to justify a priori. So we stick to the original model and still expect a correct prediction of the parameters since these extra losses are almost invisible in the linear scale that we choose for the fitting procedure.
The parameters obtained from this measurement allows to characterize the excitation of spins coupled to the RF cavity.

\subsubsection{Optical absorption}\label{sec:absorption}

\begin{figure}[ht]
\centering
\includegraphics[width=.8\columnwidth]{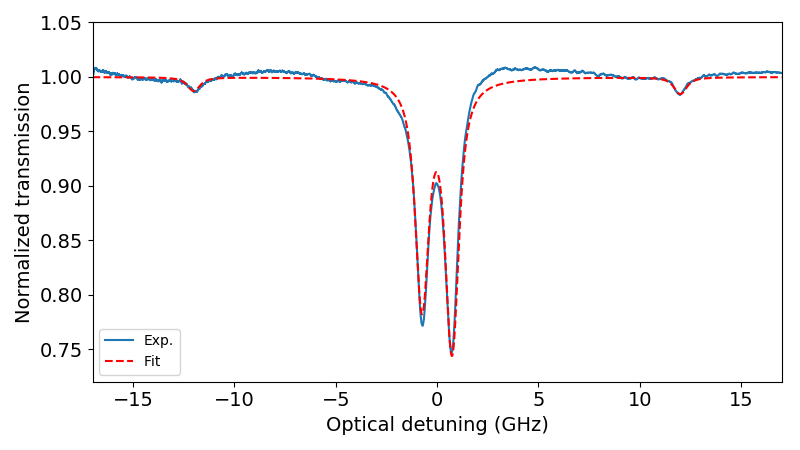} 
\caption{Absorption spectrum under a 105mT magnetic field (solid line). Four lines can be observed corresponding to level structure presented in Fig.\ref{fig:levels} (right). The optical wavelength is finely swept around the central value $\lambda_\mathrm{o}=1532.636$nm. The strongest lines correspond to the direct transitions $\ket{g_-} \rightarrow \ket{e_-} $,  $\ket{g_+} \rightarrow \ket{e_+} $ (solid arrows in Fig.\ref{fig:levels}, right) and the weakest to the crossed transitions $\ket{g_-} \rightarrow \ket{e_+} $,  $\ket{g_+} \rightarrow \ket{e_-} $ (dashed arrows in Fig.\ref{fig:levels}, right). The dashed line shows a fit to the data to extract the optical depths of the transitions (see text for details).}
\label{fig:absorption}
%/home/thierry/neel_ownCloud/neel_exchange/20220505_ErCaWO4/
\end{figure}

An equivalent analysis is conducted in the optical domain by finely sweeping our Velocity\textsuperscript{\textregistered} External Cavity Diode Laser around $\lambda_\mathrm{o}=1532.636$\,nm (absorption line position at zero magnetic field) and monitoring the crystal transmission with a photodiode in Fig.\ref{fig:absorption}. The laser propagates along $\vec{y}$ parallel to the crystalline c-axis. The polarization is along $\vec{z}$ which is also the direction of the magnetic field.

We observe four lines corresponding to the so-called direct and crossed transitions in Fig.\ref{fig:levels} (right). The two central lines are the direct ones (less than 1GHz detuning) and the crossed ones are positioned at roughly $\pm 12$\,GHz. The former are stronger that the latter, without generality because this depends on the magnetic field orientation. The imbalance between the direct transitions, namely $\ket{g_-} \rightarrow \ket{e_-} $ and  $\ket{g_+} \rightarrow \ket{e_+} $, is due to the finite temperature that partially polarize the spins into the $\ket{g_-}$ state. This is also present but much less visible on the crossed ones because of their weakness. So we model the transmission as the sum of four Voigt profiles and attribute two different absorption coefficients to the direct and crossed transitions.
A Voigt profile is difficult to justify a priori, as the broadening here is mainly inhomogeneous. However, it is a simple profile that can take into account the deformation of the transition by the presence of the isotope with hyperfine structure. The latter produces shoulders that the fit manages to capture.
We also include the finite temperature to account for the Boltzmann thermal populations. We fit data and obtain the different parameters.
The fit is visibly imperfect because of a distorted absorption baseline that we remove before fitting. The latter is not flat because of the cryostat transmission path going through many insulating uncoated windows (vacuum ports, radiation shields and cold windows).
We obtain a peak optical depth of $0.518\pm0.001$ and $0.028\pm0.001$ for the direct and crossed transitions respectively along the 4mm propagation direction of the crystal. The optical depths are normalized to a total population of one, so the finite temperature measurements exhibit a smaller absorption because the population is shared between $\ket{g_-}$ and $\ket{g_+}$. In other words, for an expected fully polarized spins ensemble, the direct and crossed transitions from $\ket{g_-}$ would have an optical depth of $0.518$ and $0.028$ respectively.

The fitted optical linewdith is $\Gamma=2\pi\times(755 \pm 25)$\,MHz (FWHM), a typical value for low concentration rare-earth samples. The value is compatible with the previously reported 1\,GHz inhomogeneous linewidth measurement in \ercawo by Sun {\it et al.} \cite{SUN2002281}.

This concludes the characterization of the spin and optical transitions independently in terms of cooperativity. The parameters extracted from both fits will be used to model the transduction efficiency in section \ref{sec:model}. The corresponding experimental results will be detailed now.

\subsection{Transduction characterization}\label{sec:transduction}
Using the detection setup sketched in Fig.\ref{fig:detection} and described in section \ref{sec:setup}, the opto-RF transduction signal appears as a $44$ MHz beating between the spin signal frequency $\omega_\mathrm{o}+\omega_\mathrm{rf}$ and the local oscillator $\omega_\mathrm{o}+\omega_\mathrm{lo}$. The photodiode is directly connected to a spectrum analyzer (Agilent E4402B, centered at 44\,MHz, 50\,kHz span, 1\,kHz resolution). The amplitude of the beatnote peaks (raw data) varies from -75\,dBm (maximum signal) to -103.5\,dBm (noise floor) corresponding to the measured voltage  $V_\mathrm{sa}$ as defined in section \ref{sec:calibration}. The data will be displayed in terms of electro-optics efficiency $\eta_\mathrm{eo}$ (in dB, left plot axis) using Eq.\eqref{eq:eom_eff} and quantum number efficiency efficiency (in dB, right plot axis) since both are simply related by Eq.\eqref{eq:eom_Q}.

We will first characterize the transduction signal as a function of the magnetic field (spin detuning) in section \ref{sec:spin_detuning} and $\omega_\mathrm{rf}$ (cavity detuning) in section \ref{sec:cav_detuning} for a given laser wavelength of $\lambda_\mathrm{o}=1532.727$\,nm corresponding to the $\ket{g_+} \rightarrow \ket{e_-} $ transition. We will finally vary the optical wavelength in section \ref{sec:wavelength} to compare the different absorption lines.

\subsubsection{Spin detuning}\label{sec:spin_detuning}

\begin{figure}[ht]
\centering
\includegraphics[width=.8\columnwidth]{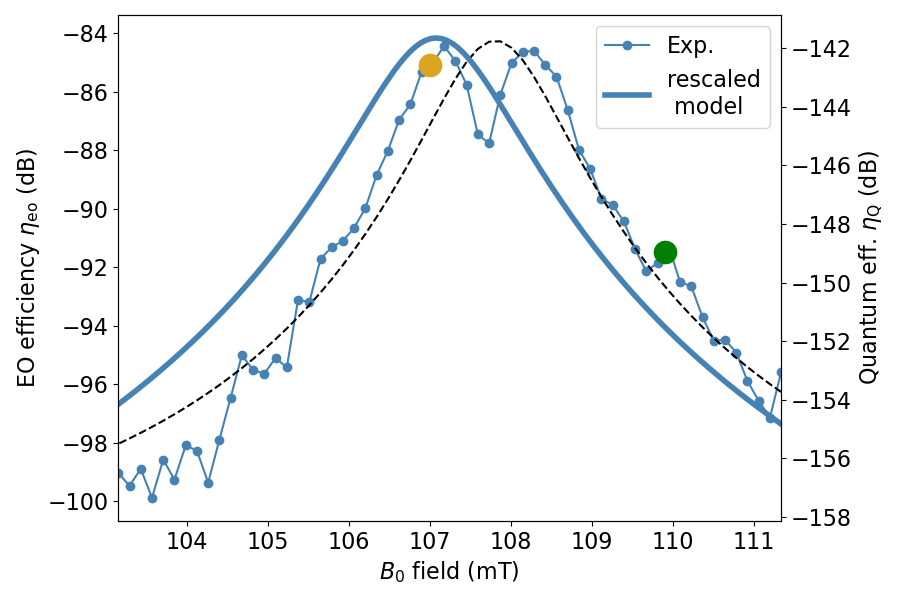} 
\caption{Transduction efficiency as a function of the spin detuning. The latter is adjusted by scanning the magnetic field across the resonance as in Fig.\ref{fig:cooperativity}. The colored circles correspond to the values $B_0=107.0$\,mT and $109.9$\,mT that will be used for a RF frequency scanning in Fig.\ref{fig:cav_detuning}. The dashed line is a lorentzian fit to the data. The left axis represents the electro-optics efficiency and the right one the quantum efficiency. Both are related by Eq.\eqref{eq:eom_Q}.
The solid line is the theoretical prediction affected by a rescaling factor as will be discussed in section \ref{sec:eom_eff}}
\label{fig:spin_detuning}
%/home/thierry/ownCloud/exp/20220505_ErCaWO4/
\end{figure}

We keep the RF and laser frequencies constant as  $\omega_\mathrm{rf}= 2 \pi \times 12.29$\,GHz and $\lambda_\mathrm{o}=1532.727$\,nm and we sweep the magnetic field to tune the spins on resonance (see Fig. \ref{fig:spin_detuning}).

The input RF power is 18.4\,dBm, the maximum power of our VNA as directly measured at low frequency ($\sim$1\,GHz). The RF power at the cavity input can be inferred from the -5.4\,dB coaxial line attenuation (extracted for the fit described in \ref{sec:RF cooperativity}) by attributing half of the losses to the input and the other half to the output, so we assume the input RF power to be reduced by -2.7\,dB at the cavity level namely $P_\mathrm{rf} = 37$\,mW or 15.7\,dBm.

The electro-optics efficiency is globally weak with a maximum of -84\,dB approximately, using Eq.\eqref{eq:eom_eff} to convert the measured beatnote into a transduction performance. Our noise detection floor is -103.5\,dB when expressed in electro-optics efficiency. A quantitative analysis will be developed in section \ref{sec:model}.

The linewidth of the beatnote $189\pm14$\,MHz obtained from the  lorentzian fit (dashed lines in Fig.\ref{fig:spin_detuning}, that include our noise detection floor) can be compared to the  spin linewidth $\gamma=2\pi\times219.14$\,MHz as measured in section \ref{sec:RF cooperativity}. An exact comparison is difficult to make, as the experimental conditions are not completely stable, as pointed out by the presence of dip at center of the experimental curve and slight shift from the spin resonance from shot to shot. The main source of instability is the presence of liquid helium at this temperature in the vicinity of the RF cavity. We have observed that, depending on the operating point of the wet cryostat, the potentially superfluid liquid helium can suddenly enter the cavity, thus slightly shifting the resonance. This is a difficult point to control during acquisition time. Apart from the linewidth, that is roughly predicted, the model is down scaled by -14\,dB to retrieve the experimental maximum as will be discussed in section \ref{sec:analysis}.

\subsubsection{Cavity detuning}\label{sec:cav_detuning}
 
\begin{figure}[ht]
\centering
\includegraphics[width=.8\columnwidth]{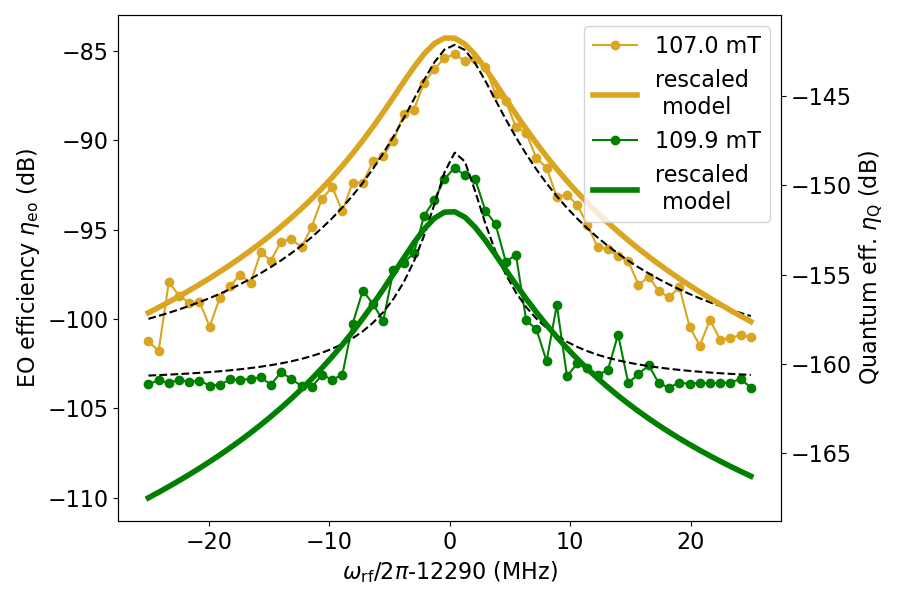} 
\caption{Transduction efficiency as a function of the cavity detuning. We vary the  $\omega_\mathrm{rf}$ around the cavity resonance at $12.290$\,GHz spanning $50$\,MHz. $B_0=107$\,mT and $109.9$\,mT correspond respectively to spins on resonance or slightly detuned from the cavity (see Fig. \ref{fig:spin_detuning}). The dashed lines are a lorentzian fit to the data. The left axis represents $\eta_\mathrm{eo}$ and the right one $\eta_\mathrm{Q}$. The dashed line is a lorentzian fit to the data. The solid lines are the expectation from the model detailed in section \ref{sec:eom_eff}}
\label{fig:cav_detuning} 
%/home/thierry/ownCloud/exp/20220505_ErCaWO4/

\end{figure}

 Under the same experimental conditions ($\lambda_\mathrm{o}=1532.727$\,nm and 18.4\,dBm RF power), we now vary the RF frequency for two values of the magnetic field ($B_0=107$\,mT and $109.9$\,mT) as plotted in Fig.\ref{fig:cav_detuning}.

One roughly retrieves the cavity linewidth on the electro-optics efficiency. The curves are well approximated by  a lorentzian fit (dashed lines in Fig.\ref{fig:cav_detuning}, that include our noise detection floor of -103.5\,dB) with $6.6\pm0.4$\,MHz and $3.6\pm0.4$\,MHz linewidths for $B_0=107$\,mT and $109.9$\,mT respectively. These values should be compared to the $\kappa_t=2\pi\times8.57$\,MHz cavity linewidth obtained from Fig.\ref{fig:cooperativity}. The second value is much narrower than $\kappa_t$, this cannot be explained so far as will be discussed in the quantitative comparison of the efficiency that will be detailed in section \ref{sec:analysis}.

\subsubsection{Wavelength dependency}\label{sec:wavelength}
\begin{figure}[ht]
\centering
\includegraphics[width=\columnwidth]{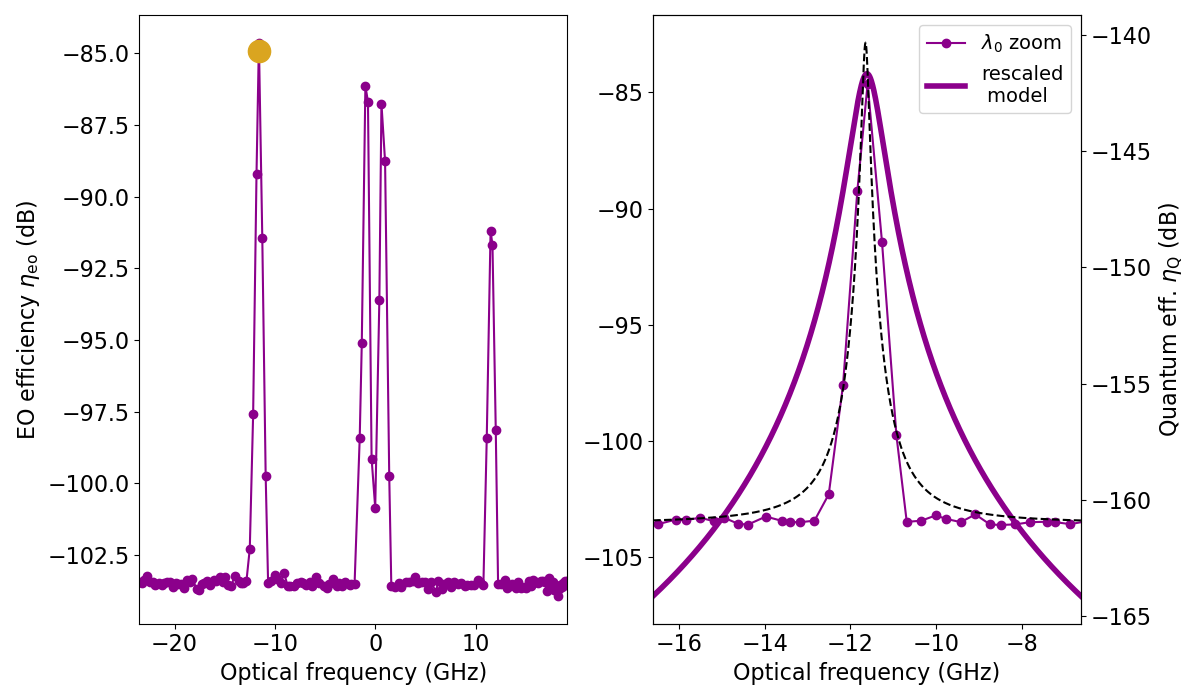} 
\caption{Transduction efficiency as a function of the optical wavelength for $B_0=107.0$\,mT. Left: broad spectrum covering the different absorption lines observed in Fig.\ref{fig:absorption}. The coloured circle correspond to a wavelength of $\lambda_\mathrm{o}=1532.727$\,nm at which other measurements have been performed. Right: Narrow scan around $\lambda_\mathrm{o}$ compared to the prediction of the model (solid line) that will be presented in section \ref{sec:eom_eff}. The dashed is a lorentzian fit to the data including our noise detection floor (linewidth $164.97\pm 0.05$\,MHz). As before, the left axis represents $\eta_\mathrm{eo}$ and the right one $\eta_\mathrm{Q}$ as in Figs.\ref{fig:spin_detuning} and \ref{fig:cav_detuning}.}
\label{fig:lambda}
%/home/thierry/ownCloud/exp/20220505_ErCaWO4/

\end{figure}

Keeping the spin RF excitation at the previously described optimal conditions, we now vary the laser excitation wavelength. The laser is finely tuned step by step ($\sim 0.3$ GHz steps). After each step, we record the laser wavelength (Burleigh WA-1000 Wavelength Meter) and acquire the transduction signal in Fig.\ref{fig:lambda} as detailed in section \ref{sec:calibration}.

The transduction spectrum in Fig.\ref{fig:lambda} should be compared to the optical absorption in Fig.\ref{fig:absorption}. We retrieve the position of the different absorption lines corresponding to the level structure of Fig.\ref{fig:levels} (right). The efficiency tends to decrease from one peak to the other when the optical frequency increases. The tendency is not reproducible from one data set to the other and may be explained by the narrowness of the lines compared to the frequency steps of $\sim 0.3$\,GHz steps which makes the measurement quite sensitive to the experimental instabilities that we discussed in section \ref{sec:spin_detuning} (presence of liquid helium  in the vicinity of the RF cavity)

In any case, the transduction efficiency does not follow the strength of the transitions. The weakest and the strongest transitions in Fig.\ref{fig:absorption} roughly give the same transduction signal. This is actually not surprising because the appearance of the transduction signal in a $\Lambda$-system connects the strength of both optical transitions involved in the non-linear process (namely $\ket{s} - \ket{e}$ and $\ket{g} - \ket{e}$, Fig.\ref{fig:levels}, left), more precisely the product of the direct and the crossed transitions strengths. The latter is generally constant within $\Lambda$-systems formed by Kramers doublets in the ground and excited states. This qualitative explanation will be discussed in the light of the theoretical model developed in the next section \ref{sec:model}. 

Finally, in Fig.\ref{fig:lambda} (right), we zoom in on one of the peaks and compare the measurement to the model. This latter predicts a transduction optical linewidth of $\Gamma=2\pi\times755$\,MHz equals to the optical absorption linewidth as in Fig.\ref{fig:absorption}. The transduction linewidth is much narrower (164.97\,MHz for the FWHM). This is a unexpected feature that will discussed in section \ref{sec:analysis}.
\section{Theoretical modeling}\label{sec:model}

The transduction process in a $\Lambda$-scheme as sketched in Fig.\ref{fig:levels} can be described by Schr\"odinger-Maxwell model that we detail in \ref{appendix:model}. We present the result in the form of an input-output solution.

\subsection{Input-output solution}

The outgoing signal field after propagation through the crystal (length $z$) reads as an integral solution for $\mathcal{E}(z)$ after a propagation $z$ Eq.\eqref{MB_transduction}:

\begin{equation}
\mathcal{E}(z)=
\frac{1}{2\delta}  \left[\exp\left(\frac{-2 i \delta \Gamma}{4 \delta \Delta - \Omega^2} \alpha_\mathcal{E}\, z\right) - 1\right] \Omega \times \mu 
\label{MB_transduction_text}
\end{equation}
The different parameters are schematically represented in Fig.\ref{fig:levels}(left).
The input-output solution \eqref{MB_transduction_text} relates the different fields represented by their Rabi frequencies: $\mu$ for RF field, $\Omega$ for the optical Raman field and $\mathcal{E}$ for the transduction signal. The experimental parameters $\Gamma$ and $\gamma$ are the optical and spin linewidths (FWHM) respectively. $\Delta$ and $\delta$ are the detunings of optical and spin transitions respectively. $  \alpha_\mathcal{E}$ is the absorption coefficient on the $\ket{g} - \ket{e}$ transition, or in other words, $\alpha_\mathcal{E}\, z$ is the optical cooperativity (optical depth).

In order to define the electro-optics efficiency, one has to compare the intensity of the optical field $\Omega$ and the generated transduction signal $\mathcal{E}$ at the output of the medium. We choose to express the fields in units of Rabi frequencies which is a natural choice to write the light-matter interaction Hamiltonian Eq.\eqref{bloch3}. Nevertheless, the comparison of the different fields intensities is then indirect and requires the introduction of the transition dipole moments. This will be discussed specifically in section \ref{sec:eom_eff}. One may be also surprised by the predominance of  the optical cooperativity  on the $\ket{g} - \ket{e}$ transition $\alpha_\mathcal{E}\, z$, because one expects a symmetric dependency with respect to the fields involved in the non-linear mixing process  $\mu$ , $\Omega$ and $\mathcal{E}$. This contradiction is only apparent because of the choice of units as will be made explicit in defining the electro-optics and quantum efficiencies in sections \ref{sec:eom_eff} and \ref{sec:eta_eff}.

In the meantime, one can already discuss and simplify the expression \eqref{MB_transduction_text}. The right-hand side term $\Omega \times \mu$ represents the lowest order frequency mixing process that generates the transduction signal from the optical and RF fields. Strictly speaking, the transduction can be seen as a sum-frequency generation, namely a second order nonlinear optical-RF process as already noted in \cite{Lambert}. The prefactor $\displaystyle \frac{2\delta \Gamma}{4 \delta \Delta - \Omega^2}$ potentially introduces high-order contributions from $ \Omega^2$,  $\Omega^4$ ... if written as a Taylor expansion so this is not a second order nonlinear process anymore. It should be nevertheless noted that the intensity $\Omega^2$ is in practice smaller than $\delta \Delta$. $\Delta$ is indeed of the order of the optical inhomogeneous linewidth, experimentally much larger than $\Omega$ \cite{kaplyanskii_spectroscopy, liu2006spectroscopic}. This further simplifies the analysis.

Additionally, one can consider the low absorption limit $\alpha_\mathcal{E}\, z \ll 1$, keeping only the first order term in the expansion of $\left[ 1-\exp\left(\cdots \alpha_\mathcal{E}\, z\right)\right]$

\begin{equation}
\mathcal{E}(z)=
 \left[ \frac{- i \Gamma}{4 \delta \Delta} \alpha_\mathcal{E}\, z \right] \Omega \times \mu \label{MB_transduction_lowOD}
\end{equation}

This simplified expression will be employed for the rest of the paper, and can be used to write the electro-optics and quantum efficiencies in sections \ref{sec:eom_eff} and \ref{sec:eta_eff}.

\subsection{Electro-optics efficiency}\label{sec:eom_eff}
\subsubsection{Definition}
As previously discussed, we compare the intensity of the input optical field $\Omega$ and the generated signal $\mathcal{E}$ to define electro-optics efficiency. The Rabi frequencies $\Omega$ and $\mathcal{E}$ are related to the electric field amplitudes $E_\Omega$ and $E_\mathcal{E}$ by $\Omega=\displaystyle\frac{\mathcal{P}_\Omega E_\Omega}{\hbar} $ and $ \mathcal{E}=\displaystyle\frac{\mathcal{P}_\mathcal{E} E_\mathcal{E}}{\hbar} $ respectively, where $\mathcal{P}_\Omega$ and $\mathcal{P}_\mathcal{E}$ are the transition dipoles of the optical transitions for the Raman and signal fields respectively.

The electro-optics efficiency is then $\eta_\mathrm{eo}=\displaystyle \frac{\left|E_\mathcal{E}\right|^2}{\left| E_\Omega\right|^2}$. This can be obtained directly from Eq.\eqref{MB_transduction_lowOD} by introducing the ratio of the transitions dipole moments (sometimes called the branching ratio) or equivalently $\displaystyle \sqrt{\frac{\alpha_\mathcal{E}}{\alpha_\Omega}}$ introducing the absorption coefficients of both transitions, leading to

\begin{equation}
\eta_\mathrm{eo}=
 \left| \frac{ \Gamma}{4 \delta \Delta}\right|^2 \alpha_\mathcal{E} z \times \alpha_\Omega z  \times \left|\mu\right|^2 \label{eta_eo}
\end{equation}

The formula involves spectroscopic parameters (absorption coefficients and linewidths) deduced from the experimental measurements on one side (see sections \ref{sec:RF cooperativity} and \ref{sec:absorption}) and the Rabi frequency of RF field $\mu$ on the other side. The expression is more symmetric since it involves equally the cooperativities $\alpha_\mathcal{E} z$ and $\alpha_\Omega z$ on the both optical transitions.
When defining the electro-optics efficiency, essentially comparing $\Omega$ and $\mathcal{E}$, unsurprisingly, the RF power $\left|\mu\right|^2$ appears as a parameter. 

In practice, the absorption of the optical transitions need to be deduced from experimental measurements and cannot be calculated ab initio, because the optical electric-dipole excitation is forbidden to the first order relying on the so-called forced electric-dipole and/or allowed magnetic-dipole (as for erbium) transitions \cite{liu2006spectroscopic}.

Eq.\eqref{eta_eo} involves the product $\alpha_\mathcal{E} \times \alpha_\Omega$ so the transduction efficiency does not depend on which transition is pumped with $\Omega$  (direct or crossed in Fig.\ref{fig:levels}). In other words, if a weak transition is pumped with $\Omega$ then a strong transition will generate $\mathcal{E}$ or the other way around, then the resulting efficiency will be the same for both cases since the product of strengths will be involved. That is the reason why in Fig.\ref{fig:lambda} the efficiency does not follow the relative strength of the excited transitions.

The RF Rabi frequency $\mu$ can be directly deduced a priori from the g-factors (magnetic dipole of the spin) and the value of the AC magnetic field inside the resonator (intracavity RF power). Unlike optical transitions, the spin transition is fully magnetic-dipole, so $\left|\mu\right|^2$ can be deduced from the RF cavity parameters. The well-known calculation of the AC magnetic field distribution in a rectangular resonator is recalled in \ref{appendix:AC_field}.

The RF Rabi frequency follows the resonance of the cavity given by the reflection coefficient Eq.\eqref{eq:S11}. As detailed in \ref{appendix:AC_field}, we finally obtain the expression \eqref{eq_A:mu} in the low spin cooperativity limit:

\begin{equation}
\mu =\frac{4 g_x}{\hbar} \sqrt{\mu_0}  \frac{a}{\sqrt{d^2+a^2}}  \sqrt{\frac{P_\mathrm{rf} }{V}} \sqrt{\frac{ 2 \kappa_c}{4\left(\omega_\mathrm{rf}-\omega_c\right)^2 + \kappa_t^2}} \label{eq:mu}
\end{equation}

where $g_x=g_\perp$ is $g$-factor along the $x$ direction of the RF field. $a$, $d$ and $V$ are the resonator dimensions and volume,  $P_\mathrm{rf}$ is the input resonator power, as defined in \ref{appendix:AC_field}.

\subsubsection{Finite temperature and population distribution}\label{sec:pop_distrib}

Moving away from the ideal structure discussed in Fig.\ref{fig:levels} (left), namely all the atomic population is in the level $\ket{g}$ (zero temperature), it is important to consider the imperfect polarization of the spin states as in our case at a temperature of few kelvins. The population is distributed between $\ket{g}$ and $\ket{s}$, written $\rho_\mathrm{gg}$ and $\rho_\mathrm{ss}$ respectively. There is no population in the excited state because of the huge energy gap.

To include the population distribution in the theoretical model, we would have to introduce the density matrix formalism. This is a tedious calculation that can be found in several texbooks \cite{Rand, Berman} and in the early derivation of Wong {\it et al.} for the Raman heterodyne signal \cite{PhysRevB.28.4993}. This gives an intuitive result at the end. Indeed, the imperfect polarization can be introduced  {\it by-hand} in the simplified formula \eqref{MB_transduction_lowOD}. The population difference  $\rho_\mathrm{gg}-\rho_\mathrm{ss}$ actually scales the RF spin coupling term, or in other words, for a perfect population balance there is no interaction with the spin. This annihilation would correspond to a perfect balance between the absorption and stimulated emission on the $\ket{g} - \ket{s}$ transition. Starting from the structure in Fig.\ref{fig:levels} (left), one could alternatively consider the reverse situation with $\ket{s}$ fully populated and show that both schemes would interfere destructively leading to the annihilation of the opto-RF process when both levels are populated. This interpretation can be formally justified using the density matrix, but for the present analysis, we can stay at the intuitive level and rescale $\mu \rightarrow \left( \rho_\mathrm{gg}-\rho_\mathrm{ss} \right) \mu$. This readily affects the efficiency as 

\begin{equation}\eta_\mathrm{eo}=
 \left| \frac{ \Gamma}{4 \delta \Delta }\right|^2 \alpha_\mathcal{E} z \times \alpha_\Omega z  \times \left|\mu\right|^2  \left( \rho_\mathrm{gg}-\rho_\mathrm{ss} \right)^2 \label{eta_eo_pop}
\end{equation}

Beyond the rescaling of the efficiency, the absorption terms $\alpha_\mathcal{E}$ and $\alpha_\Omega$ should be handled with precaution: $\alpha_\mathcal{E}$ ($\alpha_\Omega$ resp.) is the absorption coefficient on the $\ket{g} - \ket{e}$ transition ($\ket{g} - \ket{s}$ resp.) if $\ket{g}$ is fully populated ($\ket{s}$ resp.).

The absorption spectrum as in Fig.\ref{fig:absorption} actually measures $\alpha_\mathcal{E}$  and $\alpha_\Omega$ and already accounts for the current population at a given temperature in the states $\ket{g}$ and $\ket{s}$ by assuming a thermal population proportional to  $\rho_\mathrm{gg} = \displaystyle \frac{1}{1+\eT}$ and $\rho_\mathrm{ss} = \displaystyle \frac{\eT}{1+\eT}$. At our working temperature of 2.5\,K, the population imbalance is then  $\displaystyle \left( \rho_\mathrm{gg}-\rho_\mathrm{ss} \right)=0.12$. This numerical value will be used in Eq.\eqref{eta_eo_pop}.

%In practice, having measured optical depths of $\alpha_\mathcal{E} z \rho_\mathrm{gg} =0.44$ and $\alpha_\Omega z \rho_\mathrm{ss} = 0.016$ for the direct and crossed transitions respectively at 2.5K we infer  $\alpha_\mathcal{E} z=0.78$ and $\alpha_\Omega z=0.04$ that can feed \eqref{eta_eo_pop} with  $\left( \rho_\mathrm{gg}-\rho_\mathrm{ss} \right)=0.12$.

In conclusion, the electro-optics efficiency Eq.\eqref{eta_eo_pop} which includes the RF Rabi frequency Eq.\eqref{eq:mu} only contains already measured parameters that can be used to compare the experimental results with the previously derived model as we will discuss in section \ref{sec:discussion}.

\subsection{Quantum number efficiency}\label{sec:eta_eff}

\subsubsection{Definition}

As compared to the electro-optics efficiency defined in section \ref{sec:eom_eff}, it is important to properly describe the quantum efficiency. The latter is the main figure of merit for a quantum transducer. Again, Eq.\eqref{MB_transduction_lowOD} characterizes the mixing process of the optical pump $\Omega$ with the RF field $\mu$ (and vice-versa) to generate the optical signal $\mathcal{E}$. As previously explained, the electro-optics efficiency essentially compares the intensities of the signal $\mathcal{E}$ and the pump $\Omega$. The quantum efficiency compares different quantities, namely the number of photons in the optical signal $\mathcal{E}$ and the RF field $\mu$. From this point of view, the optical pump $\Omega$ is just a classical parameter.

The Rabi frequencies of the optical field $\mathcal{E}$ and the RF $\mu$  are related to the quantized fields, called $\mathcal{A}$ and $\mathcal{B}$ in the following, by the coupling constant $g_\mathcal{E}$ and $g_\mu$  respectively in the sense of the Jaynes–Cummings model. So we have $\displaystyle \mathcal{E} =  g_\mathcal{E}\, \mathcal{A}$ and $\displaystyle \mu =  g_\mu\, \mathcal{B}$ respectively.

In Eq.\eqref{MB_transduction_lowOD}, the term $\displaystyle \frac{\Gamma}{4 \delta \Delta}  \Omega$ can be taken as a parameter, but it should also be noted that $\alpha_\mathcal{E}$ also contains a contribution from the coupling rate $g_\mathcal{E}$ between light and atoms. Both are related by the following standard expressions, for the coupling constant

\begin{equation}
g_\mathcal{E}= d_\mathcal{E} \sqrt{\frac{\omega_o}{2 \hbar \epsilon_0 V_\mathcal{E}}} \label{eq:g_E}
\end{equation}
where $d_\mathcal{E}$ is the transition dipole and $V_\mathcal{E}$ the quantization volume whose choice will be discussed extensively in section \ref{sec:quantization}, and for the absorption coefficient

\begin{equation}
\alpha_\mathcal{E}=  \frac{\mathcal{N} \omega_o {d_\mathcal{E}}^2}{\hbar c \epsilon_0 \Gamma}  \label{eq:alpha_E}
\end{equation}
where $\mathcal{N}$ is the atomic density. So $\alpha_\mathcal{E}$ can be alternatively expressed as a function of $g_\mathcal{E}$ with Eq.\eqref{eq:g_E}

\begin{equation}
\alpha_\mathcal{E}=  \frac{2 \mathcal{N} V_\mathcal{E} {g_\mathcal{E}}^2}{c \Gamma}
\end{equation}

This allows us to express Eq.\eqref{MB_transduction_lowOD} in terms of quantized fields $\mathcal{A}$ and $\mathcal{B}$ as 

\begin{equation}
\mathcal{A}=
 \left[ \frac{- i \Omega}{4 \delta \Delta} \right] \, \frac{2 \mathcal{N} V_\mathcal{E} z}{c}\, g_\mathcal{E} g_\mu \,  \mathcal{B} \label{eq:aVSb}
\end{equation}

Since the optical and RF fields play an equivalent role, the expression can be made more symmetric by introducing the cooperativities that we used as key parameters to characterize the experimental data. In cavity quantum electrodynamics (cavity QED), the spin cooperativity is defined as

\begin{equation}
C_\mu=\frac{g_\mu^2 N_\mu}{\gamma \kappa_c}\label{eq:c_mu}
\end{equation}
where $N_\mu=\mathcal{N} V_\mathrm{crystal}$ is the number of excited spins ($V_\mathrm{crystal}$ is the crystal volume) \cite{Julsgaard_PhysRevA.85.013844}. Usual cavity QED expressions can be used without ambiguity for the RF spin coupling, but this is not the case for the optical excitation because there is no optical resonator in our approach. The quantization of the optical field in free-space and the definition of the corresponding cooperativity should then be handled with precaution as we will see now.

\subsubsection{Choice of quantization volume}\label{sec:quantization}

Our experimental choice, namely a resonant RF cavity but free-space optical pass, may be a source of confusion when considering the quantum efficiency by comparing the number of photons.  There is no ambiguity when defining the RF photon volume because it is contained in the RF cavity. 

On the contrary, the quantization volume for the optical field in free-space deserves a discussion. The quantization volume is composed of the transverse dimension of the optical photon that is essentially imposed by the beam size (cross-section $S_\mathcal{E}$) and the spatial extension of the wavepacket (pulse duration). Nevertheless, the optical and RF photons should have the same bandwidth and then the same pulse duration. As a consequence the quantization volume of the optical photons depends on the RF photons volume. With this choice, we impose for the optical cooperatitiy to read as 

\begin{equation}
C_\mathcal{E}=\frac{g_\mathcal{E}^2 N_\mathcal{E}}{\Gamma \kappa_c}\label{eq:c_E}
\end{equation}
$N_\mathcal{E}= \mathcal{N} S_\mathcal{E} z$ is the number of atoms in the optical path (cross-section $S_\mathcal{E}$ and length $z$). Again, we impose the optical quantization volume to depend on the RF cavity linewidth as
\begin{equation}
V_\mathcal{E}=S_\mathcal{E} \frac{c}{2 \kappa_c}
\end{equation}
where $\displaystyle \frac{c}{2 \kappa_c}$ is the photon spatial extension in free-space. This choice may appear arbitrary and surprising because $N_\mathcal{E}\neq \mathcal{N} V_\mathcal{E}$ for example. But this is driven by the physics of the conversion process and has the advantage to give consistent results at the end, somehow extending the cavity QED model to free-space approaches, that can be explored experimentally.

With this choice, one can alternatively express the optical cooperativity $C_\mathcal{E}$ defined by Eq.\eqref{eq:c_E} using the absorption coefficient  \eqref{eq:alpha_E} and the coupling constant \eqref{eq:g_E} as
\begin{equation}
C_\mathcal{E}=\alpha_\mathcal{E} z
\end{equation}
Without surprise, cooperativity and optical depth are the same in free-space, as noted early in the context of quantum memories \cite{Gorshkov}. This result is quite intuitive, thus supporting our choice of quantization volume.

\subsubsection{Expression as function of the cooperativities}

Eq. \eqref{eq:aVSb} can now be simplified by introducing the cooperativities defined by Eqs.\eqref{eq:c_mu} and \eqref{eq:c_E}.
\begin{equation}
\mathcal{A}=
 \left[ \frac{- i \Omega  \sqrt{\Gamma \gamma} }{4 \delta \Delta} \right]\,\sqrt{\frac{S_\mathcal{E}z}{V_\mathrm{crystal}}} \sqrt{C_\mathcal{E}} \sqrt{C_\mu} \cdot \mathcal{B} \label{eq:aVSb_C}
\end{equation}

The quantum number efficiency $\eta_\mathrm{Q}$ compares the photon numbers in  the quantized fields $\mathcal{A}$ and $\mathcal{B}$ leading to the expression

\begin{equation}\eta_\mathrm{Q}=
 \left| \frac{ \Omega  \sqrt{\Gamma \gamma}}{4 \delta \Delta}\right|^2 \, {C_\mathcal{E}} {C_\mu} \, \frac{S_\mathcal{E}z}{V_\mathrm{crystal}} \label{eta_Q}
\end{equation}

Eq.\eqref{eta_Q} should be compared to its equivalent electro-optics efficiency expression \eqref{eta_eo}. In a similar manner, if we account for an incomplete polarization in the ground state, the efficiency is rescaled by the factor $ \left( \rho_\mathrm{gg}-\rho_\mathrm{ss} \right)^2$ as in Eq.\eqref{eta_eo_pop}.

\subsection{Discussion}\label{sec:discussion}

Eqs.\eqref{eta_eo_pop} and \eqref{eta_Q} describe the same physical process with different points of view. The quantum efficiency \eqref{eta_Q} shows the symmetric role of the laser and RF fields cooperativities as opposed to Eq.\eqref{eta_eo_pop} where the spin collective coupling is somehow hidden in the definition of the Rabi frequency. Our derivation of Eq.\eqref{eta_Q} leads to the same expression of the cavity QED approach, when both fields are enhanced in resonators \cite{Lambert}. In section \ref{sec:eta_eff}, we simply illustrate the experimental comparison between $\eta_\mathrm{eo}$ and $\eta_\mathrm{Q}$ than can be done directly with Eq.\eqref{eq:eom_Q}. Nevertheless, we have shown that the well-established Schr\"odinger-Maxwell description leads to the same expression that can be derived from the cavity QED formalism. This is important in our case since the optical field is propagating in free-space (single-pass). In any case, Eq.\eqref{eta_Q} shows the prevalence of the cooperativities and the intensity of the pump field $ \Omega$ as key parameters that should drive the process understanding and the experimental design.

As discussed in section \ref{sec:eom_eff}, Eq.\eqref{eta_eo_pop} allows us to predict the efficiency by evaluating the experimental parameters. This is sufficient to compare the data with the model. Even if the expression \eqref{eta_Q} involves different parameters (that could be measured independently), it derives from the same expression. So for a given set, Eqs.\eqref{eta_eo_pop} and \eqref{eta_Q} are proportional and related by Eq.\eqref{eq:eom_Q}. In Figs. \ref{fig:spin_detuning}, \ref{fig:cav_detuning} and \ref{fig:lambda}, we use two efficiency axes, left and right, for Eqs.\eqref{eta_eo_pop} and \eqref{eta_Q} respectively. They simply differ by 57.5\,dB.

The discrepancy between the experimental data and the model can now be discussed specifically.

\section{Analysis}\label{sec:analysis}

In  Figs. \ref{fig:spin_detuning}, \ref{fig:cav_detuning} and \ref{fig:lambda}, the model only qualitatively reproduces the data. Indeed, to roughly reproduce the observed efficiencies, we systematically downscale the model by 14\,dBm (in the three Figs. \ref{fig:spin_detuning}, \ref{fig:cav_detuning} and \ref{fig:lambda}). More precisely, the experimental efficiency $\eta_\mathrm{eo}$ peaks typically at -84\,dBm in the optimal conditions (on-resonance fields) where the model predicts -70.2\,dBm.

Concerning the RF excitation, we note that the linewidths, when we vary the spin detuning in Fig.\ref{fig:spin_detuning} or the cavity detuning in Fig.\ref{fig:cav_detuning}, are roughly reproduced when the measured signal largely exceeds our noise level. Only the 14\,dBm discrepancy remains.

Concerning the optical linewidth of the transduction signal in Fig.\ref{fig:lambda}, we find a FWHM of 164.97\,MHz, that is much narrower than the absorption peaks $\Gamma=2\pi\times755$\,MHz in Fig.\ref{fig:lambda}. In the end, the observed value of 164.97\,MHz  is closer to the spin linewidth $\gamma=2\pi\times219.14$\,MHz measured in \ref{sec:RF cooperativity}. This points to a possible correlation between the spin and optical transitions, the effect of which has still to be assessed \cite{correlation_spin}. Even if possible correlations are not taken into account by our model, we must remain cautious about their possible impact. The spin linewidth measured independently is close to that measured on the transduction signal, unlike the optical width, which shows a large difference. Strong correlations should affect both proportionately. Additionally, the  spin linewidth is convincingly explained by the inhomogeneity of the magnetic field, i.e. by a spatial gradient. Conversely, we observe no significant broadening of the optical transition by the magnetic field. If correlations exist, they are undoubtedly weak, and can hardly explain an efficiency discrepancy of 14dBm.

The fact that we obtain a transduction feature narrower than the inhomogeneous absorption profile may also indicate that the mixing process actually addresses a small subset of ions in the overall inhomogeneous linewidth, comparable to the spectral hole burning mechanism, without reaching the homogeneous linewidth though ($<$MHz). This may partially explain the discrepancy with the model, even if we use a low laser intensity (moderate power and large beam diameter). We haven't noticed any saturation of the transduction process when the laser power is varied. It would be nevertheless interesting to include spectral hole burning and more generally optical pumping dynamics in our model. This is not the case so far because we assume in \ref{sec:perturb} that the population stays in the ground state (so-called perturbative regime), or at minimum that the population distribution is steady (see section \ref{sec:pop_distrib}). This would require further theoretical modeling and would for sure introduce another level of non-linearity with the optical pumping intensity (noted $\Omega^2$) acting not only on the coherence as in Eqs.\eqref{bloch_P} and \eqref{bloch_S}, but also on the population as an optical pumping mechanism.

We conclude with a last remark to further explain the limitations of the theoretical model and the discrepancy with the experiment. We have briefly discussed the nature of optical transitions in section \ref{sec:eom_eff} driven by $\Omega$ and $\mathcal{E}$, forced electric or magnetic dipole, both of which are possible in the case of erbium. In terms of absorption, there is no distinction. This is not the case for the transduction process. The branches of the $\Lambda$-system must in fact be of the same nature $\Omega$ and $\mathcal{E}$ without imposing any special constraint on the spin transition driven by $\mu$, as judiciously noted in \cite{Faraon_PhysRevB.104.054111, rochman2023microwave}. A fictitious pathological case, where one branch of the $\Lambda$-system would be purely magnetic dipole and the other purely electrical dipole, prohibits the non-linear mixing process. The interaction Hamiltonian introduced in the \ref{appendix:model} here assumes couplings of the same nature for the transitions, generally electric dipole for the sake of simplicity, but an equivalent model would give the same result for a magnetic dipole coupling \cite{kiruluta2006field}. A measurement of absorption solely cannot determine the nature of the transitions, and can only roughly predict the efficiency of the transduction process.
This uncertainty should be minimized in our case, since the magnetic field is perpendicular to the crystal c-axis. In the opposite situation, field parallel to the c-axis, the cross transitions are for example be forbidden in {Nd$^{3+}$:YVO$_4$, which shares many similarities with \ercawo (scheelite crystalline structure and magnetic dipole allowed transition) \cite{AFZELIUS20101566}. In this case, there is no $\Lambda$-system of magnetic dipole nature. However, a rigorous evaluation can only be made through a detailed analysis of the crystal field and the nature of the wave functions in the levels involved following  \cite{AFZELIUS20101566} or \cite{Faraon_PhysRevB.104.054111}.

%Additionally, we observe in Fig.\ref{fig:lambda} that the transduction varies from one line to the other by $\sim$6dBm. This is not expected even if this should not follow the relative strength of the lines neither as discussed in \ref{sec:eom_eff}. Instead, we expect it to be the same for the different lines. This observation clearly goes beyond our model.

\section{Conclusion}\label{sec:conclusion}
We observed and carefully characterized the opto-RF transduction process in \ercawo using the resonant excitations of the spin and optical transitions in a $\Lambda$-system.
Our work can alternatively be seen as the quantitative extension of the Raman heterodyne detection technique, since we carefully model the intensity of the beating signal.
We have decided to operate a simplified setup, employing a rectangular RF resonator to enhance the spin driving and a single-pass laser beam to excite the optical transition. This allows to extract the different experimental parameters independently that can feed our Schr\"odinger-Maxwell model.

 We reach an electro-optics efficiency of -84\,dB for 15.7\,dBm RF power, corresponding to a quantum efficiency of -142\,dB for 0.4\,dBm optical power. We carefully define both quantities that describe the same process from different points of view. The model convincingly predicts the different spin-related linewidths but fails to reproduce the optical linewidth and overestimates the efficiencies. We bring out a systematic discrepancy and downscale the model by 14\,dBm to fit data. The origin of this discrepancy is not understood. In the context of quantum transduction, where efficiency matters, we hope our analysis will stimulate further studies, both theoretical and experimental, involving other materials.
 
Increasing the efficiency to reach 100\% would require a series of improvements that are well-described in the literature. We review them to replace our work in this context. The question here is to gain orders of magnitude compared to our proof of principle, which is limited to -142\,dB. The configuration we have chosen (optical single-pass) can be optimized by using \eryso for example, whose bare cooperativity (absorption) remains larger and the transitions narrower, with a direct impact on the efficiency Eq.\eqref{eta_Q}. A higher optical power can then be used to achieve typically -90\,dB  \cite{PhysRevA.100.033807}. Another order of magnitude can be gained by fully polarizing the spins at reduced temperature (10-20\,mK), which will in any case be necessary for the interconnection with superconducting qubits for example. There's still a long way to go to reach a unit efficiency.

The use of an optical cavity then offers a typical gain of four orders of magnitude compared with single-pass configuration, as the cooperativity follows proportionally the finesse of the optical resonator \cite{PhysRevA.100.033807}. Finally, the obvious way to perfect the job is to use fully concentrated materials \cite{PhysRevA.99.063830}. Little studied in optics, they are particularly attractive compared with diluted samples (10-100\,ppm), since the concentration and therefore the cooperativity increase by 4 or 5 orders of magnitude. At very low temperatures, they can even exhibit magnetic ordering, further narrowing the magnon resonances, whose sharpness increases the efficiency\cite{doi:10.1126/science.1221878}. Their study under coupled optical and RF excitations clearly opens up a new field for rare-earth doped crystals.

\section*{Acknowledgements}
%We thank O. Arcizet for lending the extended cavity diode laser used for the measurements and the technical help of the team QuantECA on cryogenics.
%

The authors acknowledge support from the French National Research Agency (ANR) through the projects MIRESPIN (ANR-19-CE47-0011) and MARS (ANR-20-CE92-0041).
\bibliographystyle{unsrt}{}
\bibliography{transduction_bib}{}

\begin{thebibliography}{10}

\bibitem{Andrews2014}
R.~W. Andrews, R.~W. Peterson, T.~P. Purdy, K.~Cicak, R.~W. Simmonds, C.~A.
  Regal, and K.~W. Lehnert.
\newblock Bidirectional and efficient conversion between microwave and optical
  light.
\newblock {\em Nature Physics}, 10(4):321–326, March 2014.

\bibitem{Vainsencher}
Amit Vainsencher, K.~J. Satzinger, G.~A. Peairs, and A.~N. Cleland.
\newblock {Bi-directional conversion between microwave and optical frequencies
  in a piezoelectric optomechanical device}.
\newblock {\em Applied Physics Letters}, 109(3):033107, 07 2016.

\bibitem{PhysRevApplied.14.061001}
G.A. Peairs, M.-H. Chou, A.~Bienfait, H.-S. Chang, C.R. Conner, \'E. Dumur,
  J.~Grebel, R.G. Povey, E.~\ifmmode~\mbox{\c{S}}\else \c{S}\fi{}ahin, K.J.
  Satzinger, Y.P. Zhong, and A.N. Cleland.
\newblock Continuous and time-domain coherent signal conversion between optical
  and microwave frequencies.
\newblock {\em Phys. Rev. Appl.}, 14:061001, Dec 2020.

\bibitem{Mirhosseini2020}
Mohammad Mirhosseini, Alp Sipahigil, Mahmoud Kalaee, and Oskar Painter.
\newblock Superconducting qubit to optical photon transduction.
\newblock {\em Nature}, 588(7839):599–603, December 2020.

\bibitem{Delaney2022}
R.~D. Delaney, M.~D. Urmey, S.~Mittal, B.~M. Brubaker, J.~M. Kindem, P.~S.
  Burns, C.~A. Regal, and K.~W. Lehnert.
\newblock Superconducting-qubit readout via low-backaction electro-optic
  transduction.
\newblock {\em Nature}, 606(7914):489–493, June 2022.

\bibitem{PRXQuantum.1.020315}
William Hease, Alfredo Rueda, Rishabh Sahu, Matthias Wulf, Georg Arnold,
  Harald~G.L. Schwefel, and Johannes~M. Fink.
\newblock Bidirectional electro-optic wavelength conversion in the quantum
  ground state.
\newblock {\em PRX Quantum}, 1:020315, Nov 2020.

\bibitem{Sahu2022}
Rishabh Sahu, William Hease, Alfredo Rueda, Georg Arnold, Liu Qiu, and
  Johannes~M. Fink.
\newblock Quantum-enabled operation of a microwave-optical interface.
\newblock {\em Nature Communications}, 13(1), March 2022.

\bibitem{Sahu2023}
R.~Sahu, L.~Qiu, W.~Hease, G.~Arnold, Y.~Minoguchi, P.~Rabl, and J.~M. Fink.
\newblock Entangling microwaves with light.
\newblock {\em Science}, 380(6646):718–721, May 2023.

\bibitem{arnold2023alloptical}
Georg Arnold, Thomas Werner, Rishabh Sahu, Lucky~N. Kapoor, Liu Qiu, and
  Johannes~M. Fink.
\newblock All-optical single-shot readout of a superconducting qubit, 2023.

\bibitem{Lambert}
Nicholas~J. Lambert, Alfredo Rueda, Florian Sedlmeir, and Harald G.~L.
  Schwefel.
\newblock Coherent conversion between microwave and optical photons—an
  overview of physical implementations.
\newblock {\em Advanced Quantum Technologies}, 3(1):1900077, 2020.

\bibitem{Lauk_2020}
Nikolai Lauk, Neil Sinclair, Shabir Barzanjeh, Jacob~P Covey, Mark Saffman,
  Maria Spiropulu, and Christoph Simon.
\newblock Perspectives on quantum transduction.
\newblock {\em Quantum Science and Technology}, 5(2):020501, mar 2020.

\bibitem{Han:21}
Xu~Han, Wei Fu, Chang-Ling Zou, Liang Jiang, and Hong~X. Tang.
\newblock Microwave-optical quantum frequency conversion.
\newblock {\em Optica}, 8(8):1050--1064, Aug 2021.

\bibitem{PhysRevLett.113.063603}
Christopher O'Brien, Nikolai Lauk, Susanne Blum, Giovanna Morigi, and Michael
  Fleischhauer.
\newblock Interfacing superconducting qubits and telecom photons via a
  rare-earth-doped crystal.
\newblock {\em Phys. Rev. Lett.}, 113:063603, Aug 2014.

\bibitem{PhysRevLett.113.203601}
Lewis~A. Williamson, Yu-Hui Chen, and Jevon~J. Longdell.
\newblock Magneto-optic modulator with unit quantum efficiency.
\newblock {\em Phys. Rev. Lett.}, 113:203601, Nov 2014.

\bibitem{PhysRevA.99.063830}
Jonathan~R. Everts, Matthew~C. Berrington, Rose~L. Ahlefeldt, and Jevon~J.
  Longdell.
\newblock Microwave to optical photon conversion via fully concentrated
  rare-earth-ion crystals.
\newblock {\em Phys. Rev. A}, 99:063830, Jun 2019.

\bibitem{10.1063/1.1674609}
D.~E. Wortman.
\newblock {Optical Spectrum of Triply Ionized Erbium in Calcium Tungstate}.
\newblock {\em The Journal of Chemical Physics}, 54(1):314--321, 09 2003.

\bibitem{Faure1996-ep}
N~Faure, C~Borel, M~Couchaud, G~Basset, R~Templier, and C~Wyon.
\newblock Optical properties and laser performance of neodymium doped
  scheelites {CaWO4} and {NaGd(WO4)2}.
\newblock {\em Appl. Phys. B}, 63(6):593--598, December 1996.

\bibitem{CHANELIERE201877}
Thierry Chaneli{\`e}re, Gabriel H{\'e}tet, and Nicolas Sangouard.
\newblock Chapter two - quantum optical memory protocols in atomic ensembles.
\newblock volume~67 of {\em Advances In Atomic, Molecular, and Optical
  Physics}, pages 77 -- 150. Academic Press, 2018.

\bibitem{antipin1968paramagnetic}
AA~Antipin, AN~Katyshev, IN~Kurkin, and L~Ya Shekun.
\newblock Paramagnetic resonance and spin-lattice relaxation of {Er}3+ and
  {Tb}3+ ions in crystal lattice of {CaWO}4.
\newblock {\em Soviet physics solid state, USSR}, 10(2):468--+, 1968.

\bibitem{MASON1968260}
D.R. Mason and C.~Kikuchi.
\newblock Paramagnetic resonance of erbium in {CaWO4}.
\newblock {\em Physics Letters A}, 28(4):260--261, 1968.

\bibitem{mims1968phase}
WB~Mims.
\newblock Phase memory in electron spin echoes, lattice relaxation effects in
  {CaWO4}: {Er},{Ce},{Mn}.
\newblock {\em Physical Review}, 168(2):370, 1968.

\bibitem{mr-1-315-2020}
S.~Probst, G.~Zhang, M.~Ran\v{c}i\'c, V.~Ranjan, M.~Le~Dantec, Z.~Zhang,
  B.~Albanese, A.~Doll, R.~B. Liu, J.~Morton, T.~Chaneli\`ere, P.~Goldner,
  D.~Vion, D.~Esteve, and P.~Bertet.
\newblock Hyperfine spectroscopy in a quantum-limited spectrometer.
\newblock {\em Magnetic Resonance}, 1(2):315--330, 2020.

\bibitem{rochman2023microwave}
Jake Rochman, Tian Xie, John~G Bartholomew, KC~Schwab, and Andrei Faraon.
\newblock Microwave-to-optical transduction with erbium ions coupled to planar
  photonic and superconducting resonators.
\newblock {\em Nature Communications}, 14(1):1153, 2023.

\bibitem{bartholomew2020chip}
John~G Bartholomew, Jake Rochman, Tian Xie, Jonathan~M Kindem, Andrei Ruskuc,
  Ioana Craiciu, Mi~Lei, and Andrei Faraon.
\newblock On-chip coherent microwave-to-optical transduction mediated by
  ytterbium in {YVO}4.
\newblock {\em Nature communications}, 11(1):3266, 2020.

\bibitem{Photonic_Integrated_QM}
Zong-Quan Zhou, Chao Liu, Chuan-Feng Li, Guang-Can Guo, Daniel Oblak, Mi~Lei,
  Andrei Faraon, Margherita Mazzera, and Hugues de~Riedmatten.
\newblock Photonic integrated quantum memory in rare-earth doped solids.
\newblock {\em Laser \& Photonics Reviews}, 17(10):2300257, 2023.

\bibitem{PhysRevA.92.062313}
Xavier Fernandez-Gonzalvo, Yu-Hui Chen, Chunming Yin, Sven Rogge, and Jevon~J.
  Longdell.
\newblock Coherent frequency up-conversion of microwaves to the optical
  telecommunications band in an {Er:YSO} crystal.
\newblock {\em Phys. Rev. A}, 92:062313, Dec 2015.

\bibitem{PhysRevLett.50.993}
J.~Mlynek, N.~C. Wong, R.~G. DeVoe, E.~S. Kintzer, and R.~G. Brewer.
\newblock Raman heterodyne detection of nuclear magnetic resonance.
\newblock {\em Phys. Rev. Lett.}, 50:993--996, Mar 1983.

\bibitem{PhysRevB.31.6947}
M.~Mitsunaga, E.~S. Kintzer, and R.~G. Brewer.
\newblock Raman heterodyne interference: Observations and analytic theory.
\newblock {\em Phys. Rev. B}, 31:6947--6957, Jun 1985.

\bibitem{PhysRevB.28.4993}
N.~C. Wong, E.~S. Kintzer, J.~Mlynek, R.~G. DeVoe, and R.~G. Brewer.
\newblock Raman heterodyne detection of nuclear magnetic resonance.
\newblock {\em Phys. Rev. B}, 28:4993--5010, Nov 1983.

\bibitem{growth1}
S.~K. Arora and B.~Chudasama.
\newblock Crystallization and optical properties of cawo4 and srwo4.
\newblock {\em Crystal Research and Technology}, 41(11):1089--1095, 2006.

\bibitem{growth2}
Francesco Cornacchia, Alessandra Toncelli, Mauro Tonelli, Elena Favilla,
  Kirill~A. Subbotin, Valerii~A. Smirnov, Denis~A. Lis, and Evgenii~V.
  Zharikov.
\newblock {Growth and spectroscopic characterization of
  {Er}$^{3+}$:{CaWO}$_4$}.
\newblock {\em Journal of Applied Physics}, 101(12):123113, 06 2007.

\bibitem{kunkel}
Nathalie Kunkel and Philippe Goldner.
\newblock Recent advances in rare earth doped inorganic crystalline materials
  for quantum information processing.
\newblock {\em Zeitschrift für anorganische und allgemeine Chemie},
  644(2):66--76, 2018.

\bibitem{Ferrenti2020-xh}
Austin~M Ferrenti, Nathalie~P de~Leon, Jeff~D Thompson, and Robert~J Cava.
\newblock Identifying candidate hosts for quantum defects via data mining.
\newblock {\em Npj Comput. Mater.}, 6(1), August 2020.

\bibitem{bertaina2007rare}
Sylvain Bertaina, Serge Gambarelli, Alexandra Tkachuk, IN~Kurkin, Boris Malkin,
  Anatole Stepanov, and Bernard Barbara.
\newblock Rare-earth solid-state qubits.
\newblock {\em Nature nanotechnology}, 2(1):39--42, 2007.

\bibitem{dantec}
Marianne~Le Dantec, Miloš Rančić, Sen Lin, Eric Billaud, Vishal Ranjan,
  Daniel Flanigan, Sylvain Bertaina, Thierry Chanelière, Philippe Goldner,
  Andreas Erb, Ren~Bao Liu, Daniel Estève, Denis Vion, Emmanuel Flurin, and
  Patrice Bertet.
\newblock Twenty-three–millisecond electron spin coherence of erbium ions in
  a natural-abundance crystal.
\newblock {\em Science Advances}, 7(51):eabj9786, 2021.

\bibitem{Ourari2023-vc}
Salim Ourari, {\L}ukasz Dusanowski, Sebastian~P Horvath, Mehmet~T Uysal,
  Christopher~M Phenicie, Paul Stevenson, Mouktik Raha, Songtao Chen, Robert~J
  Cava, Nathalie~P de~Leon, and Jeff~D Thompson.
\newblock Indistinguishable telecom band photons from a single {Er} ion in the
  solid state.
\newblock {\em Nature}, 620(7976):977--981, August 2023.

\bibitem{Billaud_PhysRevLett.131.100804}
E.~Billaud, L.~Balembois, M.~Le~Dantec, M.~Ran\ifmmode \check{c}\else
  \v{c}\fi{}i\ifmmode~\acute{c}\else \'{c}\fi{}, E.~Albertinale, S.~Bertaina,
  T.~Chaneli\`ere, P.~Goldner, D.~Est\`eve, D.~Vion, P.~Bertet, and E.~Flurin.
\newblock Microwave fluorescence detection of spin echoes.
\newblock {\em Phys. Rev. Lett.}, 131:100804, Sep 2023.

\bibitem{wang:hal-03960036}
Zhiren Wang, L{\'e}o Balembois, Milos Ran{\v c}i{\'c}, Eric Billaud,
  Marianne~Le Dantec, Alban Ferrier, Philippe Goldner, Sylvain Bertaina,
  Thierry Chaneli{\`e}re, Daniel Est{\`e}ve, Denis Vion, Patrice Bertet, and
  Emmanuel Flurin.
\newblock {Single electron-spin-resonance detection by microwave photon
  counting}.
\newblock {\em {Nature}}, 619(7969):276--281, 2023.

\bibitem{pozar1990microwave}
D.M. Pozar.
\newblock {\em Microwave Engineering}.
\newblock Addison-Wesley series in electrical and computer engineering.
  Addison-Wesley, 1990.

\bibitem{Diniz_PhysRevA.84.063810}
I.~Diniz, S.~Portolan, R.~Ferreira, J.~M. G\'erard, P.~Bertet, and
  A.~Auff\`eves.
\newblock Strongly coupling a cavity to inhomogeneous ensembles of emitters:
  Potential for long-lived solid-state quantum memories.
\newblock {\em Phys. Rev. A}, 84:063810, Dec 2011.

\bibitem{Julsgaard_PhysRevA.85.013844}
B.~Julsgaard and K.~M\o{}lmer.
\newblock Reflectivity and transmissivity of a cavity coupled to two-level
  systems: Coherence properties and the influence of phase decay.
\newblock {\em Phys. Rev. A}, 85:013844, Jan 2012.

\bibitem{ledantec:tel-03579857}
Marianne Le~Dantec.
\newblock {\em {Electron spin dynamics of erbium ions in scheelite crystals,
  probed with superconducting resonators at millikelvin temperatures}}.
\newblock Theses, {Universit{\'e} Paris-Saclay}, January 2022.

\bibitem{enrique_optical_1971}
Bernal~G. Enrique.
\newblock Optical {Spectrum} and {Magnetic} {Properties} of {Er}3+ in
  {CaWO}$_4$.
\newblock {\em The Journal of Chemical Physics}, 55(5):2538--2549, September
  1971.

\bibitem{SUN2002281}
Y~Sun, C.W Thiel, R.L Cone, R.W Equall, and R.L Hutcheson.
\newblock Recent progress in developing new rare earth materials for hole
  burning and coherent transient applications.
\newblock {\em Journal of Luminescence}, 98(1):281--287, 2002.
\newblock Proceedings of the Seventh International Meeting on Hole Burning,
  Single Molecules and Related Spectroscopies: Science and Applications.

\bibitem{kaplyanskii_spectroscopy}
AA~Kaplyanskii and RM~McFarlane.
\newblock {\em Spectroscopy of Crystals Containing Rare Earth Ions}.
\newblock Elsevier, 1987.

\bibitem{liu2006spectroscopic}
Guokui Liu and Bernard Jacquier.
\newblock {\em Spectroscopic properties of rare earths in optical materials},
  volume~83.
\newblock Springer Science \& Business Media, 2006.

\bibitem{Rand}
Stephen~C. Rand.
\newblock {\em {Lectures on Light: Nonlinear and Quantum Optics using the
  Density Matrix}}.
\newblock Oxford University Press, 06 2016.

\bibitem{Berman}
Paul~R. Berman and Vladimir~S. Malinovsky.
\newblock {\em Principles of Laser Spectroscopy and Quantum Optics}.
\newblock Princeton University Press, 2011.

\bibitem{Gorshkov}
Alexey~V. Gorshkov, Axel Andr\'e, Mikhail~D. Lukin, and Anders~S. S\o{}rensen.
\newblock Photon storage in $\ensuremath{\Lambda}$-type optically dense atomic
  media. ii. free-space model.
\newblock {\em Phys. Rev. A}, 76:033805, Sep 2007.

\bibitem{correlation_spin}
Makoto Yamaguchi and Tohru Suemoto.
\newblock Nature of the inhomogeneous broadening of optical spectra in {Eu3+}
  doped {Y2O3} crystals studied by optical-rf double resonance.
\newblock {\em Journal of the Physical Society of Japan}, 72(2):429--436, 2003.

\bibitem{Faraon_PhysRevB.104.054111}
Tian Xie, Jake Rochman, John~G. Bartholomew, Andrei Ruskuc, Jonathan~M. Kindem,
  Ioana Craiciu, Charles~W. Thiel, Rufus~L. Cone, and Andrei Faraon.
\newblock Characterization of {Er}$^{3+}$:{YVO}$_{4}$ for microwave to optical
  transduction.
\newblock {\em Phys. Rev. B}, 104:054111, Aug 2021.

\bibitem{kiruluta2006field}
Andrew~JM Kiruluta.
\newblock Field propagation phenomena in ultra high field {NMR}: A
  {Maxwell--Bloch} formulation.
\newblock {\em Journal of Magnetic Resonance}, 182(2):308--314, 2006.

\bibitem{AFZELIUS20101566}
Mikael Afzelius, Matthias~U. Staudt, Hugues {de Riedmatten}, Nicolas Gisin,
  Olivier Guillot-Noël, Philippe Goldner, Robert Marino, Pierre Porcher,
  Enrico Cavalli, and Marco Bettinelli.
\newblock Efficient optical pumping of {Zeeman} spin levels in {Nd}3+:{YVO}4.
\newblock {\em Journal of Luminescence}, 130(9):1566--1571, 2010.
\newblock Special Issue based on the Proceedings of the Tenth International
  Meeting on Hole Burning, Single Molecule, and Related Spectroscopies: Science
  and Applications (HBSM 2009) - Issue dedicated to Ivan Lorgere and Oliver
  Guillot-Noel.

\bibitem{PhysRevA.100.033807}
Xavier Fernandez-Gonzalvo, Sebastian~P. Horvath, Yu-Hui Chen, and Jevon~J.
  Longdell.
\newblock Cavity-enhanced {Raman} heterodyne spectroscopy in
  {Er}$^{3+}$:{Y}$_2${SiO}$_5$ for microwave to optical signal conversion.
\newblock {\em Phys. Rev. A}, 100:033807, Sep 2019.

\bibitem{doi:10.1126/science.1221878}
Conradin Kraemer, Neda Nikseresht, Julian~O. Piatek, Nikolay Tsyrulin,
  Bastien~Dalla Piazza, Klaus Kiefer, Bastian Klemke, Thomas~F. Rosenbaum,
  Gabriel Aeppli, Ché Gannarelli, Karel Prokes, Andrey Podlesnyak, Thierry
  Strässle, Lukas Keller, Oksana Zaharko, Karl~W. Krämer, and Henrik~M.
  Rønnow.
\newblock Dipolar antiferromagnetism and quantum criticality in {LiErF$_4$}.
\newblock {\em Science}, 336(6087):1416--1419, 2012.

\bibitem{shore2011manipulating}
B.W. Shore.
\newblock {\em Manipulating Quantum Structures Using Laser Pulses}.
\newblock Cambridge University Press, 2011.

\end{thebibliography}

\clearpage

\appendix

%\section{Detailed fields configuration}\label{appendix:detailed}
%\section{Equivalent V$_\pi$ of the transduction device}

\section{Schr\"odinger-Maxwell model}\label{appendix:model}

The dynamics of three-level atoms under field excitation (optical or RF) can be found in many textbooks. The interest for the coherent excitation of $\Lambda$-scheme has been renewed by the observation of electromagnetically induced transparency opening a wide spectrum of quantum memory protocols \cite{CHANELIERE201877}. This represents a solid background for the theoretical description of the opto-RF transduction process even if this connection has not been made explicit so far. A modal description as a generalization of cavity quantum electrodynamics with different fields is usually preferred without specific assumption on the physical system \cite{Lauk_2020, Lambert}. We choose to stick to the Schr\"odinger equation for three-level atoms as depicted in Fig.\ref{fig:levels_Bloch} even if both approaches are equivalent to some extend. 

\begin{figure}[ht]
\centering
\includegraphics[width=.5\columnwidth]{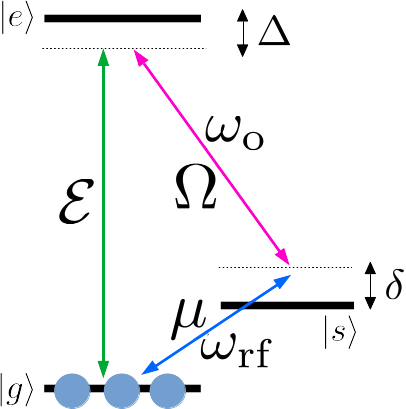} 
\caption{$\Lambda$-scheme to implement resonant transduction similar to Fig.\ref{fig:levels} with the notations for the model in \ref{sec:model}. $\mu$ is the Rabi frequency of RF field on the spin transition $\ket{g} - \ket{s}$ (frequency $\omega_\mathrm{rf}$ detuned by $\delta$). $\Omega$ is the Rabi frequency of optical field on the transition $\ket{s} - \ket{e}$ (frequency $\omega_\mathrm{o}$ detuned by $\Delta$). The transduction signal is generated at $\omega_\mathrm{o}+\omega_\mathrm{rf}$ with a Rabi frequency $\mathcal{E}$ on the $\ket{g} - \ket{e}$ transition. We first assume that the level $\ket{g}$ is solely populated (zero temperature). This assumption will be correct {\it by-hand} to account for the presence of population in $\ket{s}$ as well (finite temperature in \ref{sec:pop_distrib}).}
\label{fig:levels_Bloch}
%/home/thierry/neel_ownCloud/neel_exchange/20220505_ErCaWO4/
\end{figure}

\subsection{Schr\"odinger equation for three-level atoms in the perturbative regime}\label{sec:perturb} 
For three-level atoms, labeled $|g\rangle$,  $|e\rangle$ and  $|s\rangle$ for the ground, excited  and spin  states depicted in Fig.\ref{fig:levels_Bloch}, the rotating-wave probability amplitudes $C_g$, $C_e$ and $C_s$ respectively are governed by the time-dependent Schr\"odinger equation \cite[eq. (13.29)]{shore2011manipulating}:
\renewcommand\arraystretch{2}
\begin{align}
i \partial_t \left[
\begin{array}{c}
C_g \\
C_e \\
C_s\\
\end{array}\right]
=
\left[
\begin{array}{ccc}
0 &\displaystyle \frac{\mathcal{E}^*}{2} & \displaystyle \frac{\mu^*}{2} \\
\displaystyle\frac{\mathcal{E}}{2} & -\Delta &\displaystyle \frac{\Omega}{2} \\
\displaystyle \frac{\mu}{2} &\displaystyle \frac{\Omega^*}{2} & - \delta \\
\end{array}\right]
\left[
\begin{array}{c}
C_g \\
C_e \\
C_s\\
\end{array}\right]
\label{bloch3}\end{align}
where $\Omega(t)$ and $\mu(t)$ are the complex slowly varying envelopes of the RF and the optical (sometime called Raman) field respectively. The transduction signal $\mathcal{E}(z,t)$ potentially contains a spatial dependency in the single-pass configuration (along $z$) accounting for absorption and/or amplification though propagation. $\mathcal{E}(z,t)$ is emitted $\omega_\mathrm{o}+\omega_\mathrm{rf}$. If the spin level $|s\rangle$ is empty, the Raman field $\Omega(t)$ is not attenuated (independent of $z$). The RF field $\mu(t)$ do not depend on $z$ because the field is uniform in the crystal.

The atomic variables $C_g$, $C_e$ and $C_s$ depend on $z$ and $t$ for given detunings $\Delta$ and $\delta$. Decay terms $\Gamma$ and $\gamma$ for the optical and spin transitions can be added {\it by-hand} by introducing complex detunings for $\Delta \rightarrow \Delta - i \Gamma/2$ and $\delta \rightarrow \delta - i \gamma/2$ where $\Gamma$ and $\gamma$ are the optical and spin linewidths (FWHM) respectively.

The symmetry of the Hamiltonian \eqref{bloch3} reveals the equivalency between the optical and RF fields and the bidirectional character of the opto-RF conversion, from optics to RF and vice versa.

Going on the step further, the so-called perturbative regime assumes that the atoms stay in the ground state, $C_g \simeq 1$ to the first order, because the signal is weak. The atomic evolution (eq.\ref{bloch3}) is now only given by $C_e$ and $C_s$ that we write with $\mathcal{P}\simeq C_e$ and  $\mathcal{S}\simeq C_s$ to describe the optical (polarization $\mathcal{P}$) and spin ($\mathcal{S}$) excitations \cite[and references therein]{CHANELIERE201877}. The atoms dynamics from Eq.\eqref{bloch3} becomes:

\begin{align}
\partial_t \mathcal{P} &= i\Delta  \mathcal{P} - i \frac{\Omega}{2} \mathcal{S} - i \frac{\mathcal{E}}{2}\label{bloch_P}\\
\partial_t \mathcal{S} &=  - i \frac{\Omega^*}{2}  \mathcal{P} + i \delta \mathcal{S} - i \frac{\mu}{2} \label{bloch_S}
\end{align}

\subsection{Maxwell propagation equation}

The propagation of the signal $\mathcal{E}(z,t)$ is described by the Maxwell equation that can be simplified in the slowly varying envelope approximation \cite[eq. (21.15)]{shore2011manipulating}. This reads for an homogeneous ensemble whose linewidth is given by the decay term $\Gamma$:
\begin{equation}
\partial_z\mathcal{E}(z,t)+ \frac{1}{c}\partial_t\mathcal{E}(z,t)=
-\displaystyle {i \alpha_\mathcal{E}} \, \Gamma C_g^* C_e \label{MB_M_hom}
\end{equation}

The term $C_g^* C_e$ is the atomic coherence on the $|g\rangle \rightarrow |e\rangle$ transition directly proportional to the atomic polarization. The light coupling constant is included in the absorption coefficient $\alpha_\mathcal{E}$ on the $\ket{g} - \ket{e}$ transition (inverse of a length unit), thus the right hand side represents the macroscopic atomic polarization $\mathcal{P}$ and can be written as

\begin{equation}
\partial_z\mathcal{E}(z,t)+ \frac{1}{c}\partial_t\mathcal{E}(z,t)=
-\displaystyle{i \alpha_\mathcal{E}}\, \Gamma \mathcal{P}(t) \label{MB_M_hom_pert}
\end{equation}

The set of coupled equations \eqref{bloch_P}\&\eqref{bloch_S} and the propagation \eqref{MB_M_hom_pert} describe the generation of the transduction signal in the single optical pass configuration that we explore experimentally. We now use this model to predict the electro-optics efficiency as described in section \ref{sec:calibration}.

\subsection{Input-output solution}\label{sec:model_eff}

In the continuous pumping regime, the efficiency is given by the stationary solution of Eqs.\eqref{bloch_P}\&\eqref{bloch_S}, more precisely
\begin{equation}
\mathcal{P}= \frac{2\delta}{4 \delta \Delta - \Omega^2}\mathcal{E} + \frac{\Omega}{4 \delta \Delta - \Omega^2}\mu
\label{eq_P_stationary}
\end{equation}

The variables are now time independent. We additionally assume that the fields are real thus setting a fixed phase relation between them, in other words the fields are not frequency swept. We recall that the decay rates are hidden in the terms $\Delta \rightarrow \Delta - i \Gamma/2$ and $\delta \rightarrow \delta - i \gamma/2$ allowing a compact notation. The reader familiar with the previously mentioned electromagnetically induced transparency will recognized in the prefactor $\displaystyle \frac{2\delta}{4 \delta \Delta - \Omega^2}$ the optical susceptibility under Raman excitation which vanished (complete transparency) at $\delta=0$. The other term in $\mu$ represents the optical excitation driven by the RF field $\mu$ and the Raman field $\Omega$ (mixing process). This is the source of the transduction process, so combining Eqs.\eqref{MB_M_hom_pert} (stationary solution) and \eqref{eq_P_stationary}, we obtain

\begin{equation}
\partial_z\mathcal{E}(z)=
-i \alpha_\mathcal{E}\, \frac{2\delta \Gamma}{4 \delta \Delta - \Omega^2}\mathcal{E} - i \alpha_\mathcal{E}\, \frac{\Omega \Gamma}{4 \delta \Delta - \Omega^2}\mu %\label{MB_transduction}
\end{equation}

whose integral solution formally reads as

\begin{equation}
\mathcal{E}(z)=
\frac{1}{2\delta}  \left[\exp\left(\frac{-2 i \delta \Gamma}{4 \delta \Delta - \Omega^2} \alpha_\mathcal{E}\, z\right) - 1\right] \Omega \times \mu \label{MB_transduction}
\end{equation}

This expression in discussed in Eq.\ref{sec:model} and recalled in the text as Eq.\eqref{MB_transduction_text}.
%%%%%%%%%%%%%%%%%%%%%%%%%%%%%

\section{Intracavity RF magnetic field}\label{appendix:AC_field}

The electric and magnetic field distributions are well-known is the simple rectangular $15\times10\times20$\,mm geometry \cite[eq.(6.42)]{pozar1990microwave}. The electric field $E_y$ lies along $y$, the small dimension $b=10$\,mm. The magnetic field circulates in the $(x,z)$ plane with $B_x$ and $B_z$ components (see Fig.\ref{fig:H_field} for details). The amplitudes  are:
\begin{align}
E_y &=E_0 \sin\left( \frac{\pi x}{a}\right)  \sin\left( \frac{\pi z}{d}\right) \\
B_x &=-j \frac{E_0}{c} \frac{a}{\sqrt{d^2+a^2}}  \sin\left( \frac{\pi x}{a}\right)  \cos\left( \frac{\pi z}{d}\right) \\
B_z &=j \frac{E_0}{c} \frac{d}{\sqrt{d^2+a^2}}  \cos\left( \frac{\pi x}{a}\right)  \sin\left( \frac{\pi z}{d}\right)
\end{align}
The amplitude of the electric field $E_0$ is a scaling factor of the fields components and depends on the total energy stored in the resonator $\mathbb{E}_\mathrm{cav}$ as given by \cite[eq.(6.43)]{pozar1990microwave}:

\begin{equation}
\mathbb{E}_\mathrm{cav}=\frac{\epsilon_0 V}{8} E_0^2
\end{equation}
where $V=a b d$ is the cavity volume.

\begin{figure}[ht]
\centering
\includegraphics[width=.6\columnwidth]{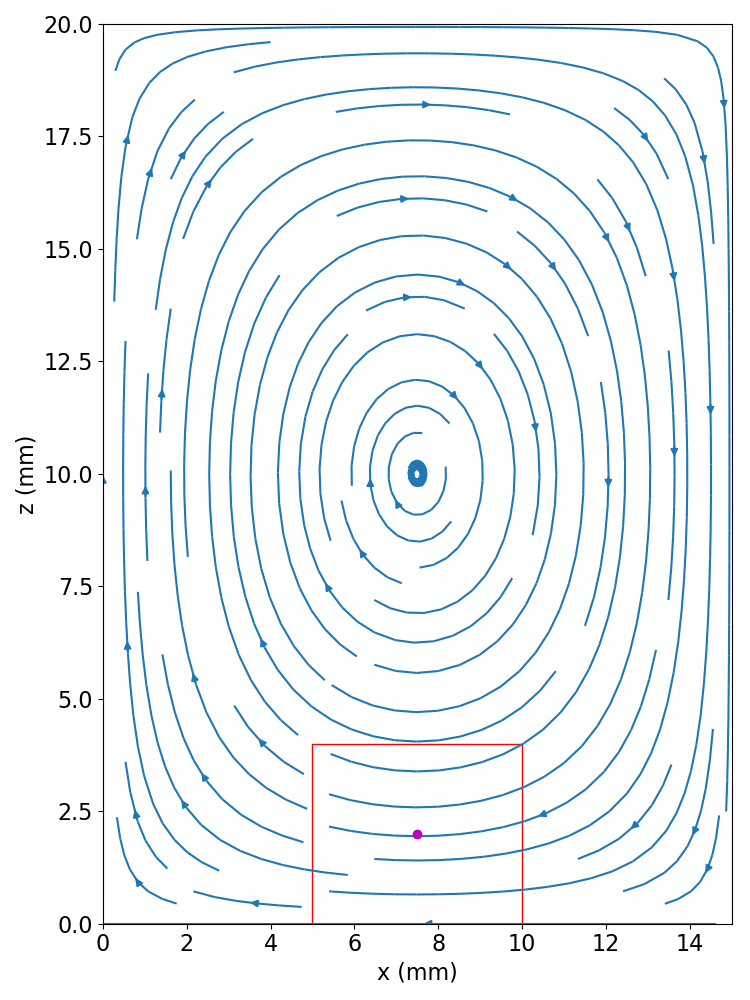} 
\caption{Magnetic field lines in the rectangular RF resonator with dimensions $a\times b\times d$=$15\times10\times20$mm along x,y,z respectively. The crystal (red square) sits at the bottom of the cavity and is crossed by the laser beam (magenta dot) propagating along y}
\label{fig:H_field}
\end{figure}

At the position of the laser schematically represented in Fig.\ref{fig:H_field} ($x_0=a/2$ and $z_0=2$\,mm), the magnetic field is uniform (small beam size compared to the cavity dimensions) and along $x$ with 
\begin{equation}
B_{x_0}(z_0)=-j \frac{E_0}{c} \frac{a}{\sqrt{d^2+a^2}}   \cos\left( \frac{\pi z_0}{d}\right)
\end{equation}
with $\displaystyle \cos\left( \frac{\pi z_0}{d}\right) \simeq 0.95$

The amplitude of the magnetic field can now be deduced from the incoming power $P_\mathrm{rf}$ and the resonator linewidth \cite[Eq.(3.16)]{ledantec:tel-03579857} and follows the resonant character of the energy stored. For an empty cavity (neglecting the spin cooperativity), we obtain

\begin{equation}
\mathbb{E}_\mathrm{cav}=\frac{\epsilon_0 V}{8} E_0^2 =\frac{4 \kappa_c}{ 4\left(\omega_\mathrm{rf}-\omega_c\right)^2 + \kappa_t^2} P_\mathrm{rf}
\end{equation}

The spin interaction can also be introduced  as in \eqref{eq:S11} where $W(\omega_\mathrm{rf})$ appears as an extra loss coefficient (spin absorption) or similarly as a rescaling of $\kappa_t$.

So the magnetic field amplitude in the crystal  is

\begin{equation}
|B_{x_0}(z_0)|= 4 \sqrt{\mu_0}  \frac{a}{\sqrt{d^2+a^2}}  \sqrt{\frac{P_\mathrm{rf} }{V}} \sqrt{\frac{ 2 \kappa_c}{4\left(\omega_\mathrm{rf}-\omega_c\right)^2 + \kappa_t^2}}\label{eq:B_x}
\end{equation}
We have dropped the term $\displaystyle \cos\left( \frac{\pi z_0}{d}\right)$ for simplicity leading to a small 5\% error. The expression can be further simplified, if needed, under critical coupling condition with $\kappa_t \approx 2\kappa_c$. Any way, the different parameters to determine to local value of the oscillating magnetic field are known for a given cavity geometry or measured from RF measurements detailed in section \ref{sec:RF cooperativity}.

The corresponding Rabi frequency, used to evaluate the electro-optics efficiency \eqref{MB_transduction_lowOD}, finally reads as 
\begin{equation}
\mu = \frac{g_x \mu_B |B_{x_0}(z_0)| }{\hbar} =  \frac{4 g_x \mu_B}{\hbar} \sqrt{\mu_0}  \frac{a}{\sqrt{d^2+a^2}}  \sqrt{\frac{P_\mathrm{rf} }{V}} \sqrt{\frac{ 2 \kappa_c}{4\left(\omega_\mathrm{rf}-\omega_c\right)^2 + \kappa_t^2}} \label{eq_A:mu}
\end{equation}
where $g_x$ is $g$-factor along the $x$ direction of the RF field. In our case, the crystalline c-axis lies along $y$ (laser propagation) so $g_x=g_\perp$, the perpendicular value of the $g$-tensor.

%%%%%%%%%%%%%%%%%%%%%%%%%%%%%

\section{Estimation of the spin cooperativity}\label{appendix:spin_cooperativity}

As introduced in section \ref{sec:RF cooperativity}, the spin cooperativity can be deduced from the cavity parameters $\gamma$, $\kappa_c$ and spin coupling strength as

\begin{equation}
C_\mu=\frac{g_\mu^2 N_\mu}{\gamma \kappa_c}
\end{equation}
where $g_\mu$ is the coupling constant and $N_\mu$ the total number of spins. $g_\mu$ can be deduced for the single-photon Rabi frequency when $\mathbb{E}_\mathrm{cav}=\hbar \omega$.

The zero-point magnetic field fluctuations using the notation of \ref{appendix:AC_field} then reads as
\begin{equation}
B_\mathrm{vac}  =2\frac{a}{\sqrt{d^2+a^2}} \sqrt{\frac{2 \hbar \omega \mu_0}{ V} }
\end{equation}
leading to the coupling strength $g_\mu= g_x \mu_B B_\mathrm{vac} /\hbar$
\begin{equation}
g_\mu =g_x \mu_B \frac{a}{\sqrt{d^2+a^2}} \sqrt{\frac{2 \omega \mu_0}{\hbar V} }\label{eq:g_0}
\end{equation}

As already noted in section \ref{sec:pop_distrib}, because of the finite temperature and the imperfect spin polarization, we have to account for the population difference  $(\rho_\mathrm{gg}-\rho_\mathrm{ss})$ that scales the spin cooperativity as well, effectively modifying the number of spins $N_\mu \rightarrow \left( \rho_\mathrm{gg}-\rho_\mathrm{ss} \right) N_\mu$. At our temperature of 2.5\,K, we have  $\displaystyle \left( \rho_\mathrm{gg}-\rho_\mathrm{ss} \right)=0.12$ (see section \ref{sec:pop_distrib}).

The cooperativity can now be evaluated for a 50\,ppm doped sample, by knowing the Ca concentration ($1.3\times10^{22}$\,at/cm$^3$) and noting that 77\% of the erbium ions (without nuclear spin) are probed by our setup, we obtain
$C_\mu=0.12$, in satisfying agreement the experimentally inferred value of 0.135034 in section \ref{sec:RF cooperativity}.

\section{Choice of the heterodyne beating frequency}\label{appendix:choice_beating}
One may be surprised by our choice  of the heterodyne beating frequency that can be in principle arbitrarily taken within the detector bandwidth.

Indeed, when we record the noise spectrum on the photodiode, we interestingly observe a minimum close to 44\,MHz that we choose as a beatnote heterodyne detection frequency as discussed in \ref{transduction_detection}.
\begin{figure}[ht]
\centering
\includegraphics[width=.6\columnwidth]{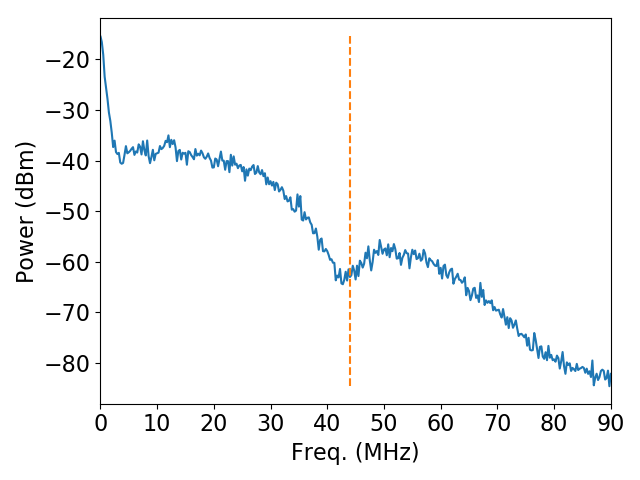} 
\caption{Noise spectrum on the photodiode (500\,kHz resolution). We mark with a vertical dashed line the minimum close to 44\,MHz that we choose as a beatnote heterodyne detection frequency for the transduction signal.}
\label{fig:Noise_spectrum}
%/home/thierry/neel_ownCloud/neel_exchange/20220505_ErCaWO4/
\end{figure}

This minimum of the laser intensity noise that our partially fibered optical setup forms a Mach-Zender interferometer for the optical carrier (see Fig.\ref{fig:detection}). The interferometer can be seen as a filter in the frequency domain, that exhibits minima for frequencies $\displaystyle \frac{c}{2L}$, $\displaystyle\frac{3c}{2L}$, $\displaystyle\frac{5c}{2L}$ ... where $L$ is the optical path length difference between the two arms. This latter is approximately $3$\,m in our case, corresponding to the observed $44$\,MHz minimum.

%\section{Density matrix derivation of the electro-optics efficiency}\label{appendix:OBE}

\end{document}